\documentclass[aps,prb,twocolumn,superscriptaddress,floatfix,longbibliography]{revtex4-2}
\usepackage{amsmath,amssymb,amsthm,bm}
\usepackage{graphicx}
\usepackage{booktabs}
\usepackage{xcolor}
\usepackage{tikz}
\usetikzlibrary{arrows.meta,positioning,shapes.geometric,fit,backgrounds,calc}
\usepackage[colorlinks=true,allcolors=blue!50!black]{hyperref}

\newtheorem{lemma}{Lemma}
\newtheorem{proposition}{Proposition}
\newtheorem{theorem}{Theorem}
\theoremstyle{remark}
\newtheorem{remark}{Remark}

\newcommand{\Tr}{\operatorname{Tr}}
\newcommand{\ee}{\mathrm{e}}
\newcommand{\ii}{\mathrm{i}}
\newcommand{\SM}{Supplemental Material~\cite{SM}}
\definecolor{lrftblue}{HTML}{2a78d6}
\definecolor{inkgray}{HTML}{6b6a64}

\begin{document}

\title{Local-resolvent formulation and assessment of linear-scaling Kohn--Sham calculations}

\author{Albert Linda}
\email{albert.hzbn@gmail.com}
\affiliation{Department of Materials Science and Engineering, Indian Institute of Technology Kanpur,
Kanpur 208016, Uttar Pradesh, India}

\date{\today}

\begin{abstract}
Kohn--Sham density-functional theory is expensive because it needs orbitals that extend over the whole system. For
insulators this can be avoided by using only the diagonal of the one-electron resolvent,
$g(\bm r,z)=\langle\bm r|(z-\hat h)^{-1}|\bm r\rangle$, at a few complex energies $z$. We call the scheme local-resolvent
density-functional theory (LRes-DFT). The density and the band energy are linear in $g$, so neither orbitals nor a
kinetic-energy functional are needed. The complex energies are $M$ equally spaced points on a circle that encloses the occupied levels, offset by half a spacing so
that no point lies on the real axis. Summing the resolvent over them is equivalent to occupying each level with
$1/(1+u^M)$, where $u$ is the level's energy measured from the circle's center in units of its radius. This is a smooth step
that is near 1 inside the circle and near 0 outside, and it stays between 0 and 1 for every $M$. If the circle moves with the
chemical potential, forces at fixed electron number are exact. Gauss--Legendre and Zolotarev rules need fewer points when
the gap is small. In three dimensions we use eigenpairs of small windows around each molecule instead of resolvents. This is
divide-and-conquer (DC) with a different occupation function, and it is as accurate as DC with Fermi--Dirac occupations. For
H$_2$ molecular crystals of up to 1024 electrons it agrees with Kohn--Sham to $2.5\times10^{-7}$~Ha per electron. On one
core it is slower than dense diagonalization below about 500 electrons and
3.5 times faster at 1024 electrons. It loses accuracy for disordered or compressed systems and
for diffuse basis functions.
\end{abstract}

\maketitle

\section{Introduction}

Density-functional theory (DFT)~\cite{HK1964,KS1965} is the standard tool for computing the energetics of materials.
In materials problems the cost of Kohn--Sham (KS) calculations is often the limiting factor. This includes generalized
stacking-fault energies of alloys~\cite{Linda2024PRB}, solute segregation at grain boundaries~\cite{Linda2026Acta},
and large charge-density data sets~\cite{Kumar2026JCIM}. The KS kinetic energy is computed from $N/2$ orbitals that extend
over the whole cell, so storage grows as $\mathcal O(N^2)$ and diagonalization or orthogonalization as $\mathcal O(N^3)$
with the number of electrons $N$. For insulators, in which the density matrix is local, this cost can be reduced to
$\mathcal O(N)$. The present work is restricted to such systems: gapped molecular solids, including disordered ones.
Metals, which are central to the applications above, require finite electronic temperature and are not treated here.

Two routes to lower cost are well established. Orbital-free DFT~\cite{WangTeter1992,WGC1999,HuangCarter2010} keeps only the
density and approximates the kinetic energy. Linear-scaling KS methods keep the exact kinetic energy and use the locality
(nearsightedness) of the density matrix~\cite{Kohn1996,ProdanKohn2005,Goedecker1999,BowlerMiyazaki2012}. They include
divide-and-conquer (DC)~\cite{Yang1991} and fragment methods~\cite{LS3DF2008}, density-matrix
minimization~\cite{LNV1993,ONETEP2005}, localized-orbital methods~\cite{MGC1993,BigDFTLS2014}, Fermi-operator and Chebyshev
expansions~\cite{Goedecker1999,WeisseKPM2006}, Krylov-subspace methods~\cite{Ozaki2006}, recursion~\cite{Haydock1972} and
stochastic DFT~\cite{BaerSDFT2013}.

A third family uses Green's functions on a complex energy contour. Korringa--Kohn--Rostoker (KKR) methods integrate on a
complex contour as a matter of course~\cite{ZellerDeutz1982}, and the LSMS, LSGF and tight-binding KKR variants compute
site-diagonal Green's functions in local zones~\cite{LSMS1995,LSGF1996,Zeller2008}. Contour-integral forms of the Fermi
operator~\cite{Goedecker1993,BaroniGiannozzi1992} and pole expansions of the Fermi--Dirac
function~\cite{LinLuYingE2009,LinMultipole2009,Moussa2016,Ozaki2007} compute the density from a few resolvents. With selected
inversion, pole expansions scale as $\mathcal O(N)$, $\mathcal O(N^{1.5})$ and $\mathcal O(N^2)$ in one, two and three
dimensions~\cite{PEXSI2013}. The same rational filters also underlie contour-integral
eigensolvers~\cite{SakuraiSugiura2003,Polizzi2009,Guettel2015,AustinTrefethen2015}. Section~\ref{sec:comparison} compares
these approaches with the present one.

The formulation used here belongs to this third family. Its starting point, the KS density as a contour integral of the
diagonal resolvent, is not new. We call the resulting scheme local-resolvent DFT (LRes-DFT). We study one simple realization in detail, with proofs and
measurements, and report what it delivers and what it does not. The paper does four things.
\begin{enumerate}
\item It gives the formulation and its limits: which quantities are linear in $g$, why the diagonal resolvent exists only
for a discretized Hamiltonian, and which operators it cannot represent.
\item It analyzes the half-step trapezoid rule on a circle (the Butterworth filter of contour-integral
eigensolvers~\cite{Guettel2015}) as a KS occupation function. The occupation $1/(1+u^M)$ always lies in $[0,1]$, the error
obeys an a priori bound, and a rigidly moving circle gives exact fixed-$N$ forces at every $M$. Gauss--Legendre and
Zolotarev rules need fewer poles, and we count how many.
\item It evaluates the occupation in local windows in three dimensions, which is a DC method. We compare it, in
self-consistent calculations of up to 1024 electrons, with DC using Fermi--Dirac occupations, with dense KS diagonalization
(the fastest of three LAPACK paths) and with a warm-started iterative KS solver.
\item It reports the cases in which the method is inaccurate.
\end{enumerate}
We checked the analytic identities with computer algebra on explicit instances. Code and data are available~\cite{SM}.

In brief, the half-step rule works as designed in one dimension. In three dimensions the method reduces to divide-and-conquer, it agrees with Kohn--Sham to about $10^{-7}$~Ha per electron for gapped molecular crystals, and it is faster than dense diagonalization only above roughly 500 electrons.

The development proceeds as follows. Section~\ref{sec:theory} gives the formulation, with proofs in
Appendix~\ref{app:proofs}. Section~\ref{sec:algorithms} describes the algorithms and Section~\ref{sec:details} the
computational details. Sections~\ref{sec:res1d} and \ref{sec:res3d} present results in one and three dimensions,
Section~\ref{sec:comparison} relates the method to other approaches, and Section~\ref{sec:limits} summarizes the limitations.

\begin{figure*}[t]
\includegraphics[width=\textwidth]{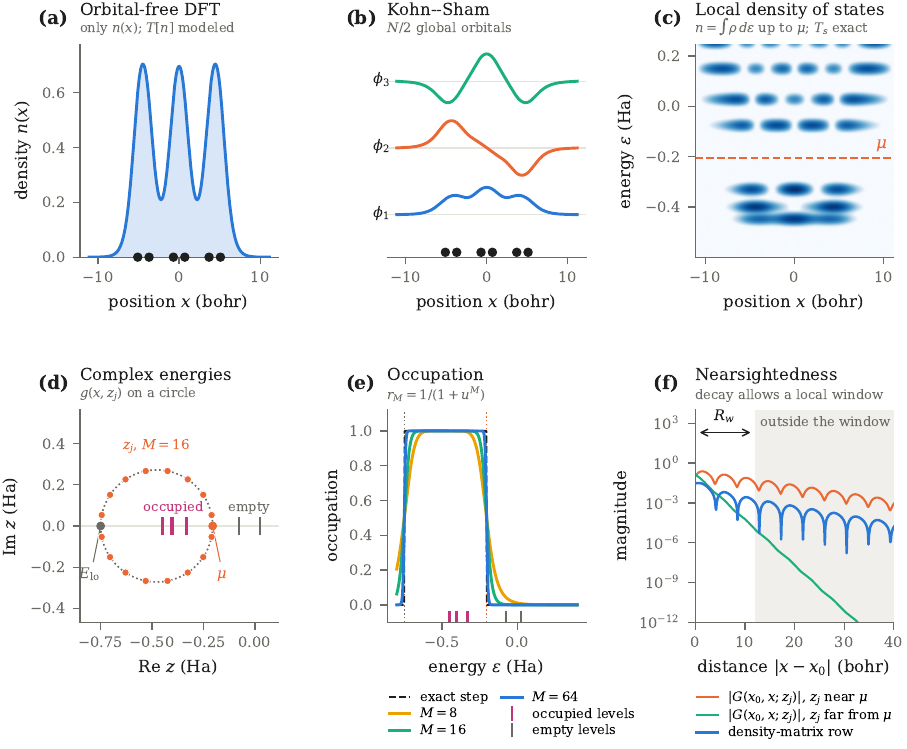}
\caption{\label{fig:concept}
Formulation (symbols are defined in Section~\ref{sec:theory}), shown for a one-dimensional (H$_2$)$_3$ chain (black dots: nuclei) in panels (a)--(e) and an (H$_2$)$_{24}$
chain in panel (f).
(a)~Orbital-free DFT keeps only the density $n(x)$ and models the kinetic energy.
(b)~KS DFT computes $N/2$ orbitals $\phi_i$ that extend over the whole system.
(c)~Local density of states $\rho(x,\varepsilon)$ (color map, logarithmic scale). The density is its integral up to the
chemical potential $\mu$, and $g(x,z)$ is its transform to complex energies.
(d)~$g$ is sampled at $M$ complex energies $z_j$ (orange) on a circle through $E_\mathrm{lo}$ and $\mu$ that encloses the
occupied levels (magenta ticks) and excludes the empty ones (gray ticks); the axes are to scale.
(e)~The half-step trapezoid rule gives the occupation $r_M=1/(1+u^M)$, which equals $1/2$ at $E_\mathrm{lo}$ and at $\mu$; the ticks at the bottom mark the occupied (magenta) and empty (gray) levels of the chain.
(f)~The resolvent $G(x_0,x;z_j)$ and the density matrix decay exponentially with distance in this gapped system, so the
density in a region can be computed from a window of radius $R_w$ (the shaded part lies outside the window).}
\end{figure*}

\section{Formulation}\label{sec:theory}

Table~\ref{tab:notation} lists the main symbols used below.

\begin{table}[t]
\caption{\label{tab:notation}Main symbols.}
\begin{ruledtabular}
\footnotesize
\begin{tabular}{ll}
Symbol & Meaning \\
\hline
$g(\bm r,z)$ & local resolvent, Equation~\eqref{eq:g} \\
$z_j$, $w_j$ & nodes and weights on the circle \\
$c$, $R$ & center and radius of the circle \\
$\mu$, $E_\mathrm{lo}$ & chemical potential; lower edge of the circle \\
$M$ & number of complex energies \\
$u$, $r_M(u)$ & $(\varepsilon-c)/R$; occupation $1/(1+u^M)$ \\
$\Delta$, $W_\mathrm{occ}$ & gap and occupied band width \\
$R_w$ & window radius (3D) \\
$N$, $N_b$ & electrons; basis functions \\
\end{tabular}
\end{ruledtabular}
\end{table}

\subsection{The local resolvent}

We consider spin-unpolarized electrons in an effective one-electron Hamiltonian $\hat h=-\tfrac12\nabla^2+v(\bm r)$, with
$v=v_\mathrm{ext}+v_\mathrm{H}[n]+v_\mathrm{xc}[n]$ the external, Hartree and exchange--correlation (xc) potentials
(Hartree atomic units). Throughout, $\hat h$ is the Hamiltonian discretized on a grid or in a basis, i.e., a finite
Hermitian matrix with eigenvalues $\varepsilon_i$ and orthonormal eigenvectors $\phi_i$. On a grid with volume element
$\Delta V$ we use orthonormal grid vectors $|\bm r\rangle$, so that $|\phi_i(\bm r)|^2$ is a probability per grid point and
$n(\bm r)=2\sum_{i\,\mathrm{occ}}|\phi_i(\bm r)|^2/\Delta V$. The local resolvent is
\begin{equation}
g(\bm r,z)=\langle\bm r|(z-\hat h)^{-1}|\bm r\rangle=\sum_i\frac{|\phi_i(\bm r)|^2}{z-\varepsilon_i},
\label{eq:g}
\end{equation}
for complex $z$ away from the eigenvalues. The eigenvectors appear here only to explain what $g$ is; the algorithms do
not compute global eigenvectors.

\begin{remark}[Continuum limit]\label{rem:continuum}
Everything below is for a discretized $\hat h$. In the continuum the diagonal of the resolvent is finite in one dimension but
not in three: the free Green's
function $G_0(\bm r,\bm r';z)=-\ee^{\ii k|\bm r-\bm r'|}/(2\pi|\bm r-\bm r'|)$, $k=\sqrt{2z}$, diverges as $\bm r'\to\bm r$,
and on a grid $g(\bm r,z)$ grows as $1/\Delta x$. The divergent term, $-1/(2\pi|\bm r-\bm r'|)$, does not depend on $z$, so it
drops out of every contour integral of $g$ over a closed contour, and the density and band energy below remain finite.
The local resolvent itself is defined only for a discretized Hamiltonian.
\end{remark}

\subsection{What is linear in $g$}

\begin{proposition}\label{prop:rep}
Let $C$ be a closed contour, traversed counterclockwise, that encloses the occupied eigenvalues and no others. Then
(a)~$n(\bm r)\Delta V=2\cdot\frac{1}{2\pi\ii}\oint_C g(\bm r,z)\,dz$, and
(b)~$E_\mathrm{band}\equiv2\sum_{i\,\mathrm{occ}}\varepsilon_i=2\cdot\frac{1}{2\pi\ii}\oint_C z\sum_{\bm r}g(\bm r,z)\,dz$.
\end{proposition}

The KS kinetic energy follows as $T_s=E_\mathrm{band}-\sum_{\bm r}v(\bm r)n(\bm r)\Delta V$. The kinetic energy $T_s$ is not linear in $g$, but it follows from $g$ without approximation through $E_\mathrm{band}$ and
$n$, and the total energy has the usual KS form,
\begin{align}
E={}&E_\mathrm{band}-\textstyle\sum_{\bm r}v_\mathrm{Hxc}(\bm r)\,n(\bm r)\,\Delta V\nonumber\\
&+E_\mathrm{H}[n]+E_\mathrm{xc}[n]+E_{nn},
\label{eq:energy}
\end{align}
with $v_\mathrm{Hxc}=v_\mathrm{H}+v_\mathrm{xc}$ and $E_{nn}$ the nuclear repulsion.

\begin{remark}[Scope]\label{rem:potential}
The diagonal $g$ on a contour that encloses the whole spectrum contains far more information than the density. For a grid
Hamiltonian with a local potential, the large-contour moment $\frac{1}{2\pi\ii}\oint_{|z|=R'}z\,g(\bm r,z)\,dz$ ($R'$ larger
than the spectral radius) equals $h_{\bm r\bm r}=t_{\bm r\bm r}+v(\bm r)$, where $t$ is the discretized kinetic operator.
This identity fails in a non-orthogonal basis and with nonlocal pseudopotentials.

Some quantities need off-diagonal elements of the density matrix: nonlocal pseudopotential energies, exact exchange and
meta-GGA kinetic-energy densities. These are not given by $g$, so the formulation as stated is limited to local potentials
and semilocal functionals.
\end{remark}

\subsection{The half-step trapezoid rule}\label{sec:rule}

We take $C$ to be a circle that encloses the occupied levels and excludes the empty ones. It passes through a point
$E_\mathrm{lo}$ below the lowest eigenvalue and through the chemical potential $\mu$, which lies in the gap. Its center and
radius are $c=(\mu+E_\mathrm{lo})/2$ and $R=(\mu-E_\mathrm{lo})/2$. We discretize it with an even number $M$ of points at
half-step angles,
\begin{equation}
z_j=c+R\zeta_j,\qquad \zeta_j=\ee^{\ii\pi(2j+1)/M},\qquad w_j=\frac{R\zeta_j}{M},
\label{eq:nodes}
\end{equation}
$j=0,\dots,M-1$. We call the $z_j$ nodes; in the pole-expansion literature they are called poles, and the cost is counted in
complex solves. The $\zeta_j$ are the roots of $\zeta^M=-1$, and $w_j$ are the trapezoid weights of
$\frac{1}{2\pi\ii}\oint dz$ (the factor $1/2\pi\ii$ is included). The nodes come in complex-conjugate pairs, so only
$M/2$ resolvents are needed for a real symmetric $\hat h$.

Summing the resolvent over the nodes can be done in closed form, which gives the occupation function directly.

\begin{lemma}[Closed form]\label{lem:closed}
$\frac1M\sum_j\zeta_j/(\zeta_j-u)=1/(1+u^M)$ for all complex $u\neq\zeta_j$. Hence
\begin{equation}
\hat P_M\equiv\sum_{j}w_j(z_j-\hat h)^{-1}=\Big[1+\Big(\frac{\hat h-c}{R}\Big)^M\Big]^{-1},
\label{eq:PM}
\end{equation}
and in a non-orthogonal basis with overlap $S>0$, $\sum_jw_j(z_jS-H)^{-1}=r_M(S^{-1}H)S^{-1}$.
\end{lemma}

The approximate density is $n_M(\bm r)\Delta V=2[\hat P_M]_{\bm r\bm r}$, and each eigenstate is occupied with weight
\begin{equation}
r_M(u)=\frac{1}{1+u^M},\qquad u=\frac{\varepsilon-c}{R}.
\end{equation}
For $M=16$, a state with $u=0.8$ receives $0.973$ and a state with $u=1.25$ receives $0.027$; for $M=64$ these become
$1-6.3\times10^{-7}$ and $6.3\times10^{-7}$. This filter is not new: it is the Butterworth filter that the half-step trapezoid rule on a circle gives in contour-integral
eigensolvers~\cite{Guettel2015,AustinTrefethen2015}. What we add is its use as the occupation function of a KS calculation,
with the consequences derived below.

Because the occupation is an explicit function, its range and its error for a single state can be bounded.

\begin{lemma}[Occupations and error per state]\label{lem:bounds}
For even $M$ and real $u$, $r_M(u)=1/[1+\exp(M\ln|u|)]$, so $0<r_M\le1$. Hence $0\preceq\hat P_M\preceq1$, and $\hat P_M$ is a valid (ensemble-representable) one-body
density matrix~\cite{Coleman1963}. If $|u|\le\rho<1$, then $1-r_M\le\rho^M$; if $|u|\ge1/\rho$, then
$r_M\le\rho^M$.
\end{lemma}

Two remarks on this rule. First, the ordinary trapezoid placement ($\zeta^M=+1$) gives $1/(1-u^M)$, which exceeds 1 inside
the circle and is negative outside; the half-step placement avoids this. Second, unlike a Fermi--Dirac function, $r_M$ is
not monotonic in $\varepsilon$: it increases for $\varepsilon<c$ and decreases for $\varepsilon>c$, so a state below
$E_\mathrm{lo}$ is occupied by less than $1/2$. No concave entropy-like function generates $r_M$, so the finite-temperature
(Mermin) minimum principle~\cite{Mermin1965} has no direct analog here. Consequently we do not claim that the energy is
minimal. We show only that it is stationary, which is what forces require.

\subsection{Stationarity and forces}\label{sec:var}

We next show that $n_M$ is the exact derivative of a functional of the potential. This gives stationarity of the energy
and exact forces.

\begin{proposition}\label{prop:var}
Let $\Omega_M[v]=\Tr W_M(\hat h[v])$ with $W_M(\varepsilon)=-2\sum_jw_j\ln(z_j-\varepsilon)$. $W_M$ is analytic for real
$\varepsilon$ because no node lies on the real axis, and real because the nodes and weights come in conjugate pairs;
$W_M'=2r_M$. For fixed $c$ and $R$, $\partial\Omega_M/\partial v(\bm r)=n_M(\bm r)\Delta V$ exactly.
\end{proposition}

This identity holds for any smooth occupation function, including the Gauss--Legendre and Zolotarev rules below. Its
consequence is that the self-consistent solution is a stationary point of
\begin{align}
G_M[v]={}&\Omega_M[v]-\textstyle\sum_{\bm r}(v-v_\mathrm{ext})\,n_M[v]\,\Delta V\nonumber\\
&+E_\mathrm{H}[n_M[v]]+E_\mathrm{xc}[n_M[v]]+E_{nn}
\label{eq:GM}
\end{align}
at fixed $c$ and $R$. Hence the Hellmann--Feynman force (the derivative of $\hat h$ at fixed density) is the exact derivative
of $G_M$ (Appendix~\ref{app:proofs}).
Here $\Omega_M$ plays the role of a grand potential and $G_M$ adds the double-counting terms, so this is a grand-canonical
statement. To conserve the electron number, the circle is translated along the energy axis together with $\mu$, with its radius
unchanged, so that $E_\mathrm{lo}$ moves with $\mu$. The functional $A_M=G_M+\mu N$ then has $N$ as a constraint:

\begin{proposition}\label{prop:rigid}
Keep $R$ fixed and set $c=\mu-R$. Then $\partial\Omega_M/\partial\mu=-N_M$, with $N_M=\Tr\,2\hat P_M$, and
$A_M=G_M+\mu N$ is stationary with respect to $v$ and $\mu$ at the self-consistent solution with $N_M=N$. The
Hellmann--Feynman force equals $-dA_M/dX$ exactly at every $M$.
\end{proposition}

If $E_\mathrm{lo}$ is held fixed instead, the radius $R$ changes with $\mu$. The fixed-$N$ energy,
Equation~\eqref{eq:energy}, is then stationary only up to terms that vanish as the occupation error vanishes. Forces are
correct only approximately at small $M$ (Section~\ref{sec:forces}).

\subsection{Error bound}

The density error of the half-step rule falls as $\rho^M$, where $\rho<1$ measures how close the worst state lies to the
circle. Here $\rho$ is the largest of three numbers. They measure how far $|u|$ of the lowest level, of the highest occupied level
and of the lowest empty level lies from the circle $|u|=1$ (occupied levels must have $|u|<1$, empty levels $|u|>1$). The
precise statement is:

\begin{theorem}\label{thm:conv}
Let $E_\mathrm{min}$ be the lowest eigenvalue, $E_\mathrm{lo}=E_\mathrm{min}-\delta$, $\Delta_-=\mu-\varepsilon_\mathrm{HOMO}$ and
$\Delta_+=\varepsilon_\mathrm{LUMO}-\mu$, where HOMO and LUMO denote the highest occupied and lowest unoccupied states, and let
\begin{equation}
\rho=\max\Big(\Big|1-\frac{\delta}{R}\Big|,\ 1-\frac{\Delta_-}{R},\ \frac{1}{1+\Delta_+/R}\Big)<1 .
\end{equation}
Then $\sum_{\bm r}|n_M-n|\Delta V\le2(N_\mathrm{occ}+N_\mathrm{near})\rho^M+2N_g2^{-M}$, where $N_\mathrm{occ}=N/2$,
$N_\mathrm{near}$ is the number of empty states with $1<|u|<2$, and $N_g$ is the dimension of $\hat h$.
\end{theorem}

Here $2R=W_\mathrm{occ}+\Delta_-+\delta$, with $W_\mathrm{occ}=\varepsilon_\mathrm{HOMO}-E_\mathrm{min}$ the occupied band
width. The rate does not depend on the largest eigenvalue of $\hat h$, which grows as $1/\Delta x^2$ on a grid; the bound
depends on $N_g$ only logarithmically, through $2N_g2^{-M}$. The bound also shows the weakness of the rule. For small gaps
$\ln(1/\rho)\approx\Delta/(2R)$, so $M$ grows as $(W_\mathrm{occ}/\Delta)\ln(1/\epsilon)$ for a target accuracy $\epsilon$,
whereas optimal rational approximations need $\mathcal O[\ln(W/\Delta)\ln(1/\epsilon)]$ poles, with $W$ the full spectral
width~\cite{LinLuYingE2009,Moussa2016,Guettel2015}. The theorem applies to the 1D implementation; for the 3D windows
there is no a priori bound, and we measure the window error instead.

\subsection{Cost}

In one dimension, a three-point finite-difference (FD) Hamiltonian is tridiagonal, and the diagonal of its resolvent can
be computed exactly in $\mathcal O(N_g)$ operations:

\begin{lemma}[Two-sided recursion]\label{lem:recursion}
For tridiagonal $A=z-\hat h$ with diagonal $a_i$ and off-diagonal $b_i$, let $L_1=a_1$, $L_i=a_i-b_{i-1}^2/L_{i-1}$ and
$U_{N_g}=a_{N_g}$, $U_i=a_i-b_i^2/U_{i+1}$. Then $[A^{-1}]_{ii}=1/(L_i+U_i-a_i)$.
\end{lemma}

This is the standard recursion for the diagonal of a tridiagonal inverse~\cite{Meurant1992}. It is also used in the recursive
Green's function method~\cite{Lake1997}, and in 1D it is equivalent to selected inversion~\cite{PEXSI2013}.

In more than one dimension we rely on nearsightedness. In a gapped system the resolvent and the density matrix decay
exponentially with distance~\cite{CombesThomas1973,Kohn1996,ProdanKohn2005}, so the density in a region can be obtained from
a local window [Figure~\ref{fig:concept}(f)]. We do not prove a bound for truncated windows. We measure their error
(Section~\ref{sec:res3d}).

\section{Algorithms}\label{sec:algorithms}

\begin{figure}[t]
\centering
\begin{tikzpicture}[
  node distance=3.0mm,
  box/.style={draw=inkgray, rounded corners=2pt, align=center, font=\scriptsize, inner sep=3pt, text width=5.6cm, fill=white},
  lr/.style={box, draw=lrftblue, fill=lrftblue!6, text width=5.3cm},
  dec/.style={draw=inkgray, diamond, aspect=2.6, align=center, font=\scriptsize, inner sep=1pt, fill=white},
  arr/.style={-{Stealth[length=2mm]}, inkgray, thick}]
\node[box] (n0) {Initial density $n_\mathrm{in}(\bm r)$};
\node[box, below=of n0] (v) {Potential $v=v_\mathrm{ext}+v_\mathrm{H}[n_\mathrm{in}]+v_\mathrm{xc}[n_\mathrm{in}]$};
\node[lr, below=8mm of v] (c1) {\textbf{1.} Circle through $E_\mathrm{lo}$ (below the spectrum) and $\mu$ (in the gap); reuse $\mu$, $E_\mathrm{lo}$ of the previous step if valid};
\node[lr, below=of c1] (c2) {\textbf{2.} For every region $B$, independently: window $W(B)$ of radius $R_w$; 1D: $g(\bm r,z_j)$ by Lemma~\ref{lem:recursion}; 3D: window eigenpairs below $c+1.5R$};
\node[lr, below=of c2] (c3) {\textbf{3.} $N(\mu)$, $E_\mathrm{band}$ in closed form; re-solve $\mu$ if $|N(\mu)-N|$ exceeds the tolerance, reusing the stored window vectors};
\node[lr, below=of c3] (c4) {\textbf{4.} $n_\mathrm{out}(\bm r)$ on the points owned by $B$};
\node[dec, below=7mm of c4] (conv) {converged?};
\node[box, below=5mm of conv] (E) {Total energy, Equation~\eqref{eq:energy}};
\node[box, text width=1.0cm, inner sep=2pt] (mix) at (3.7,0 |- conv.center) {Pulay\\ mixing};
\begin{scope}[on background layer]
\node[draw=lrftblue, dashed, rounded corners=3pt, fit=(c1)(c4), inner sep=4pt] (lrbox) {};
\end{scope}
\node[font=\scriptsize\bfseries, text=lrftblue, anchor=south west, inner sep=1pt] at (lrbox.north west) {LRes-DFT step};
\draw[arr] (n0) -- (v);
\draw[arr] (v) -- (c1);
\draw[arr] (c1) -- (c2);
\draw[arr] (c2) -- (c3);
\draw[arr] (c3) -- (c4);
\draw[arr] (c4) -- (conv);
\draw[arr] (conv) -- node[right, font=\scriptsize]{yes} (E);
\draw[arr] (conv) -- node[above, font=\scriptsize]{no} (mix);
\draw[arr] (mix.north) -- (mix.north |- v.east) -- (v.east);
\end{tikzpicture}
\caption{\label{fig:flow}
SCF cycle of LRes-DFT. Only the dashed box differs from KS DFT, where the same step diagonalizes the full Hamiltonian.
Each region is computed from its own window, so the work grows with the number of regions.}
\end{figure}
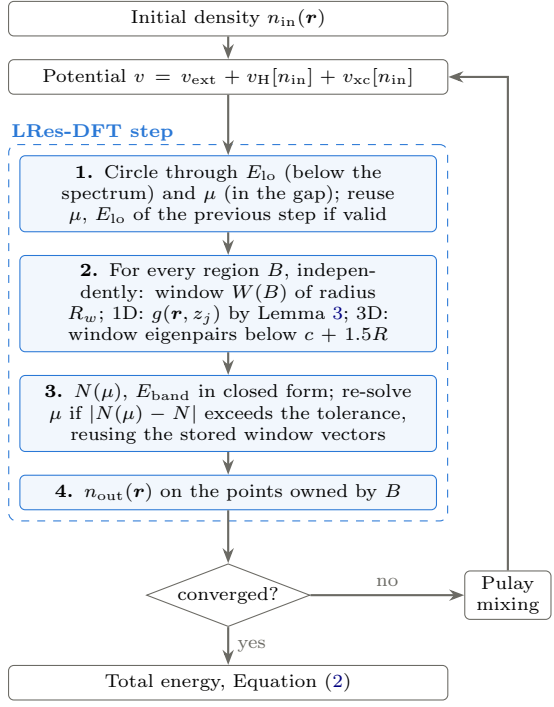

The self-consistent field (SCF) cycle is the same as in KS DFT, except for the step that turns a potential into a
density (Figure~\ref{fig:flow}). We describe this step first in one dimension, where $g$ is computed exactly by a recursion,
and then in three dimensions, where the density comes from eigenpairs of small windows.

\subsection{One dimension}

In 1D, $g(x,z_j)$ is computed at all grid points with Lemma~\ref{lem:recursion} for $M/2$ nodes, and the density follows
from Lemma~\ref{lem:closed}; no eigenvectors are formed. The chemical potential is found by a secant search on $N(\mu)$,
started from the previous SCF step. $E_\mathrm{lo}$ is the lowest eigenvalue, obtained by bisection, minus a margin
$\delta$.

\subsection{Three dimensions: local windows}\label{sec:windows}

The 3D algorithm has four ingredients: a localized basis, a division of the system into regions with windows, the density
from window eigenpairs weighted by $r_M$, and a chemical potential obtained from the electron count in closed form. We
describe them in turn.

\paragraph*{Basis.} We use localized basis functions $\chi_\mu$ tabulated on the grid, a linear combination of atomic orbitals
(LCAO), as production linear-scaling codes do~\cite{SIESTA2002,ONETEP2005}. The matrices are
$H=X^{\mathsf T}(T_\mathrm{FD}+v)X\,\Delta V$ and $S=X^{\mathsf T}X\,\Delta V$. Here $X$ holds the basis functions on the grid
and $T_\mathrm{FD}$ is the FD kinetic operator.

\paragraph*{Regions and windows.} The system is divided into regions, one per molecule. Each grid point is assigned to the
nearest molecule, that is, to the region $B$ whose Voronoi cell contains it; we say that $B$ owns these points. The density at
an owned point is computed from the window $W(B)$, which contains the basis functions of all molecules within a radius $R_w$
of the center of $B$.

\paragraph*{Density from window eigenpairs.} On each window we solve $H_W\phi_k=\lambda_kS_W\phi_k$ and set, for
$\bm r\in B$,
\begin{equation}
n(\bm r)=2\sum_kr_M\Big(\frac{\lambda_k-c}{R}\Big)|\phi_k(\bm r)|^2 .
\label{eq:window}
\end{equation}
By Lemma~\ref{lem:closed} this equals the pole sum $2\sum_jw_jg_W(\bm r,z_j)$ of the window Hamiltonian, so
Equation~\eqref{eq:window} is a DC construction~\cite{Yang1991} with the occupation $r_M$. In 3D we therefore never form the
local resolvent. Eigenpairs of a small window weighted by $r_M$ are equivalent to summing $M$ window resolvents and are about
20 times cheaper (Section~\ref{sec:res3d}). Here $M$ enters only through the occupation function.

\paragraph*{Electron number.} The electron number and band energy,
$N(\mu)=2\sum_{B,k}c_{Bk}r_M(u_{Bk})$ and $E_\mathrm{band}=2\sum_{B,k}c_{Bk}\lambda_{Bk}r_M(u_{Bk})$ with
$c_{Bk}=\sum_{\bm r\in B}|\phi_k(\bm r)|^2\Delta V$, are explicit functions of $\mu$. The chemical potential is therefore
found by solving $N(\mu)=N$ without further diagonalizations.

\paragraph*{SCF steps.} Once $\mu$ is known, only eigenpairs below $c+1.5R$ are computed, because $r_M<1.5^{-M}$ above that
energy. Each SCF step then proceeds as follows.
\begin{enumerate}
\item In the first (``cold'') step all window eigenvalues are computed, to place $\mu$ and $E_\mathrm{lo}$.
\item In later (``warm'') steps $\mu$ and $E_\mathrm{lo}$ are reused, so only the low eigenpairs are computed.
\item If the electron count must be corrected, $\mu$ is re-solved from the closed-form $N(\mu)$, and the density is rebuilt
from the stored $|\phi_k(\bm r)|^2$ on the owned points. No second eigensolver pass is needed.
\item $E_\mathrm{lo}$ is first set to the lowest window eigenvalue minus $\delta$. It then stays fixed while the lowest window
eigenvalue remains between $E_\mathrm{lo}+\delta/3$ and $E_\mathrm{lo}+2\delta$, that is, while the margin stays within a
factor of 3 of $\delta$. Otherwise the circle is re-centered: $E_\mathrm{lo}$ is reset to that eigenvalue minus $\delta$.
\end{enumerate}

\paragraph*{Why not point by point?} We first evaluated $g$ point by point on a real-space grid (spacing 0.6~bohr) with
probing vectors and Lanczos quadrature~\cite{WeisseKPM2006}. The quadrature itself was accurate, but probing needs far too
many vectors: in our test crystal the density matrix extends over about $10^3$ grid points, and with 125 probes (spacing 3~bohr) the error of the probed diagonal was 3.7 times the density itself, and with 1000 probes it was still 7\%
(\SM). This is why we use a localized basis and windows.

\begin{lemma}[Window spectra are bounded below]\label{lem:interlace}
For a symmetric pencil $(H,S)$ with $S>0$, the lowest eigenvalue of any principal sub-pencil $(H_W,S_W)$ is not lower than
the lowest eigenvalue of $(H,S)$.
\end{lemma}

The lemma follows from the Courant--Fischer characterization (Appendix~\ref{app:proofs}). Truncating the Hamiltonian
to a window therefore cannot create spurious states below the true band bottom. The choice of $E_\mathrm{lo}$ above places
the circle below every window spectrum by construction, and within $\delta$ of the true band bottom as $R_w$ grows.

\section{Computational details}\label{sec:details}

\paragraph*{One dimension.} Electrons interact through the soft-Coulomb potential $1/\sqrt{x^2+1}$ and are attracted by
nuclei through $-Z/\sqrt{(x-X)^2+1}$, a common one-dimensional model~\cite{Helbig2011}. The grid spacing is 0.1~bohr, with
6--10~bohr of vacuum and hard walls. We use the local exchange of the 1D soft-Coulomb uniform gas in the closed form
$e_x(n)=-\pi^{-2}[bI_{K_0}(b)+bK_1(b)-1]$, $b=\pi n$, $I_{K_0}(b)=\int_0^bK_0$ (Appendix~\ref{app:ex}), without correlation.
We refer to this as the exchange-only local-density approximation (LDA). Exact two-electron energies are obtained by diagonalization in the two-particle
space on the same grid. The KS reference uses the identical Hamiltonian and is solved for the occupied states with LAPACK
bisection (\texttt{stebz}) followed by inverse iteration (\texttt{stein}). This was faster than the multiple relatively robust
representations (MRRR) algorithm (\texttt{stemr}) in our tests (\SM). Unless stated otherwise $\delta=0.3$~Ha.

\paragraph*{Three dimensions.}
\begin{description}
\item[System.] H$_2$ molecular crystals of $n\times n\times n$ cubic cells (lattice constant $a=6$~bohr unless stated), with
random molecular orientations and random displacements of 0.15~bohr, in periodic boundary conditions. The HOMO--LUMO gaps are
0.29--0.41~Ha.
\item[Hamiltonian.] Nuclei are local pseudopotentials $-\mathrm{erf}(r/r_c)/r$ with $r_c=0.6$~bohr. The grid spacing is
0.6~bohr, the Hartree potential is obtained by fast Fourier transform, and we use the LDA with Slater exchange and
Perdew--Zunger correlation~\cite{PZ1981}.
\item[Basis.] Six functions per atom: $s$-type Gaussians with exponents 1.2, 0.4 and 0.15~bohr$^{-2}$ and a $p$ shell with
exponent 0.5~bohr$^{-2}$, with a smooth cutoff at 4.5~bohr. Its error relative to a converged real-space-grid calculation,
solved with the locally optimal block preconditioned conjugate gradient method (LOBPCG)~\cite{LOBPCG2001}, is 0.019~Ha per
electron for 16 electrons.
\item[KS reference.] KS and LRes-DFT use identical $H$, $S$, Hartree and xc terms. The dense KS reference solves the
generalized eigenproblem for the lowest $N/2+1$ states. We compared three LAPACK paths: \texttt{dsygvx} (bisection and
inverse iteration), \texttt{dsygvd} (divide and conquer) and Cholesky reduction with \texttt{dsyevr} (MRRR). We also compared
two ways of building the density. We use the fastest combination, \texttt{dsygvx} with the orbital sum (\SM). As a second
reference we use LOBPCG on the pencil $(H,S)$, warm-started from the previous SCF step, with a block-Jacobi preconditioner.
\item[LRes-DFT.] $M=64$, $\delta=0.3$~Ha and $R_w=12.1$~bohr unless stated.
\end{description}

\paragraph*{Timing.} Both 3D one-body solvers are written in C++ against the same OpenBLAS LAPACK, and the 1D LRes-DFT
kernel is compiled with Numba. All timings are single-threaded unless stated, on an 8-core workstation. The headline
comparison is the wall time of complete SCF calculations (Table~\ref{tab:rigorous}). Each size was run three times, with
the order of KS and LRes-DFT alternated, on an idle machine pinned to one core. The earlier density-build series of
Table~\ref{tab:cost3d} and the thread timings were taken while the machine was shared (load average about 1); for those we
report the best of two or three repetitions. In a separate series under heavier load, repeated KS timings varied by up to
20\% (\SM). Power-law exponents of the time are fitted to the four largest sizes, and those of the memory to the three
largest. SCF calculations use Pulay mixing~\cite{Pulay1980}.

\paragraph*{Use of AI tools.} The AI assistant Claude Sonnet 5.5 (Anthropic, model identifier \texttt{claude-sonnet-5-5}),
run through the Claude Code command-line tool, was used to write and debug the Python and C++ code, the automated tests, the
computer-algebra checks, the benchmark and plotting scripts, and the scripts that build the figures and tables from the stored
results. The author specified the model, the algorithms and the benchmarks, reviewed the code, and ran the calculations. The
results rest on the stored output of the calculations, which the tests and the computer-algebra checks cover. No figure was
generated by an AI image tool; all figures are plots of the stored data.

\paragraph*{Verification.} We checked the analytic identities with SymPy and 40-digit mpmath arithmetic on explicit
instances (30 checks). The checks cover Lemmas~\ref{lem:closed}--\ref{lem:interlace}, Propositions~\ref{prop:rep}--\ref{prop:rigid}, the
stationarity of $G_M$, the LDA potentials, the pseudopotential transform and the exchange closed form. Automated tests (25)
check the implementation against these identities and against dense linear algebra.

\section{Results in one dimension}\label{sec:res1d}

The 1D calculations test the pole rule itself. We check the error bound, count the complex solves needed by three pole
rules, compare LRes-DFT with KS and with exact two-electron solutions, test forces, and measure the cost.

\subsection{Error control}

\begin{figure*}[t]
\includegraphics[width=\textwidth]{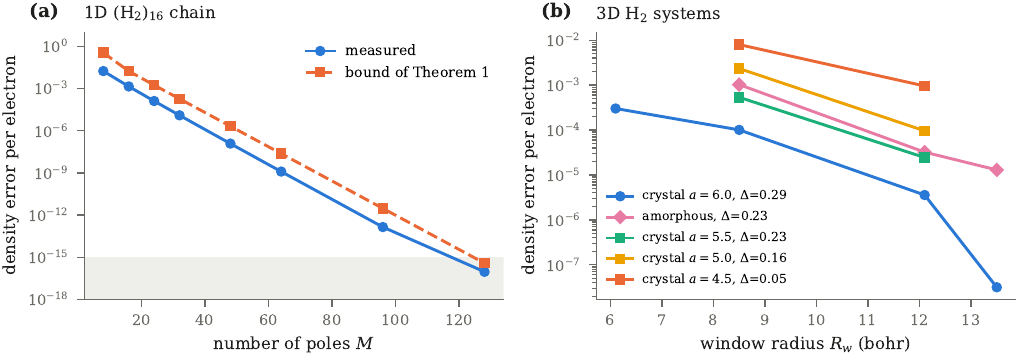}
\caption{\label{fig:conv}
Error control.
(a)~1D (H$_2$)$_{16}$ chain at the self-consistent KS potential ($\delta=0.3$~Ha): density error per electron (circles)
versus the number of poles $M$, and the bound of Theorem~\ref{thm:conv} (squares).
(b)~3D H$_2$ systems with 250--432 electrons at their self-consistent KS potentials: density error per electron versus the
window radius $R_w$ for the crystal ($a=6$~bohr), an amorphous arrangement at the same density, and compressed crystals
($a=5.5$, 5.0 and 4.5~bohr). Gaps are given in the legend.}
\end{figure*}

Figure~\ref{fig:conv}(a) shows the density error of a 1D (H$_2$)$_{16}$ chain at its self-consistent KS potential as a
function of $M$, together with the bound of Theorem~\ref{thm:conv}. The error decreases exponentially, stays below the bound,
and reaches $1.3\times10^{-9}$ per electron at $M=64$. The margin $\delta$ matters. With $M=32$ the error per electron is
$1.3\times10^{-5}$ for $\delta=0.3$~Ha and $2.1\times10^{-7}$ for $\delta=0.1$~Ha, which is about half the gap (\SM). The
results here use the conservative $\delta=0.3$~Ha; choosing $\delta\approx\Delta/2$ would reduce the required $M$.

\subsection{Comparison of pole rules}\label{sec:rules}

\begin{figure*}[t]
\includegraphics[width=\textwidth]{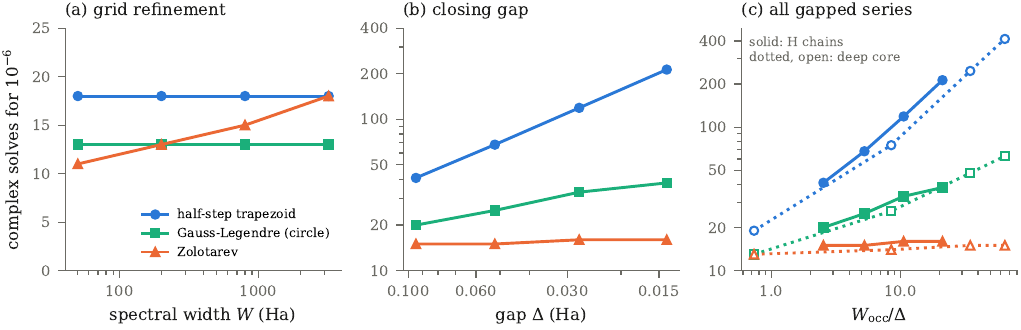}
\caption{\label{fig:rules}
Number of complex solves needed for a density error of $10^{-6}$ per electron with the half-step trapezoid rule,
Gauss--Legendre on the same circle, and Zolotarev's rational approximation. (a)~Grid refinement at fixed gap (spectral
widths 50--3200~Ha). (b)~Closing gap in uniform H chains. (c)~All gapped cases against the ratio $W_\mathrm{occ}/\Delta$ of
occupied band width to gap; solid lines and filled symbols: H chains; dotted lines and open symbols: chains with a heavier
atom.}
\end{figure*}

Table~\ref{tab:rules} and Figure~\ref{fig:rules} compare the number of complex solves needed for a density error of
$10^{-6}$ per electron. The three rules are the half-step trapezoid rule ($M/2$ solves); Gauss--Legendre in the angle on
the same circle (GL), the default of FEAST~\cite{Polizzi2009}; and Zolotarev's best rational approximation of the sign
function (the occupation step is $[1-\mathrm{sgn}(\varepsilon-\mu)]/2$, so a rational approximation of the sign gives a pole expansion)~\cite{Guettel2015,Nakatsukasa2016} on $[-W,-\Delta/2]\cup[\Delta/2,W]$, centered at $\mu$. All are evaluated
exactly on the eigendecomposition of the same self-consistent KS Hamiltonian.

We consider three series. The first is grid refinement of an (H$_2$)$_4$ chain at fixed gap. The second is uniform H$_n$
chains whose gap closes as $n$ grows. The third is an (H$_2$)$_4$ chain with one heavier atom of charge $Z$, whose deep
core state widens the occupied band; in this series the gap also decreases with $Z$.

\begin{table}[t]
\caption{\label{tab:rules}Complex solves for $10^{-6}$ density accuracy (1D) with the half-step rule (HS), Gauss--Legendre on
the circle (GL) and Zolotarev's rule (Zol.). $W$: spectral width; $\Delta$: gap; $W_\mathrm{occ}$: occupied band width (Ha).}
\begin{ruledtabular}
\footnotesize\setlength{\tabcolsep}{1.6pt}
\begin{tabular}{lrrrrrrr}
Case & $W$ & $\Delta$ & $W_\mathrm{occ}$ & $W_\mathrm{occ}/\Delta$ & HS & GL & Zol. \\
\hline
grid, $\Delta x=0.2$ & 50 & 0.232 & 0.14 & 0.6 & 18 & 13 & 11 \\
grid, $\Delta x=0.1$ & 200 & 0.233 & 0.14 & 0.6 & 18 & 13 & 13 \\
grid, $\Delta x=0.05$ & 800 & 0.233 & 0.14 & 0.6 & 18 & 13 & 15 \\
grid, $\Delta x=0.025$ & 3200 & 0.233 & 0.14 & 0.6 & 18 & 13 & 18 \\
H$_{10}$ chain & 200 & 0.095 & 0.24 & 2.5 & 41 & 20 & 15 \\
H$_{20}$ chain & 200 & 0.052 & 0.27 & 5.3 & 68 & 25 & 15 \\
H$_{40}$ chain & 200 & 0.027 & 0.29 & 10.6 & 119 & 33 & 16 \\
H$_{80}$ chain & 200 & 0.014 & 0.30 & 21.2 & 213 & 38 & 16 \\
core, $Z=2$ & 200 & 0.229 & 0.17 & 0.7 & 19 & 13 & 13 \\
core, $Z=4$ & 200 & 0.107 & 0.91 & 8.5 & 75 & 26 & 14 \\
core, $Z=6$ & 200 & 0.053 & 1.86 & 34.8 & 247 & 48 & 15 \\
core, $Z=8$ & 200 & 0.045 & 2.88 & 64.5 & 413 & 63 & 15 \\

\end{tabular}
\end{ruledtabular}
\end{table}

Under grid refinement the half-step rule needs 18 solves at every spacing, while the spectral width grows from 50 to
3200~Ha. Gauss--Legendre on the same circle also does not depend on the grid and needs 13. Zolotarev's rule grows slowly
(11 to 18), as expected from its $\ln(W/\Delta)$ dependence. As the gap closes, the half-step rule becomes much more
expensive: 41 to 213 solves, against 20 to 38 for Gauss--Legendre and 15 to 16 for Zolotarev. Plotted against
$W_\mathrm{occ}/\Delta$ [Figure~\ref{fig:rules}(c)], the H chains and the chains with a heavier atom fall on one curve, as
Theorem~\ref{thm:conv} predicts. The cost of the half-step rule is therefore set by the ratio of occupied band width to gap,
not by the deep state as such. The advantage of the half-step rule is that its occupations lie exactly in $[0,1]$
(Lemma~\ref{lem:bounds}). At the accuracy considered, the Gauss--Legendre and Zolotarev occupations deviate from $[0,1]$ by
up to $2.4\times10^{-5}$ and $1.7\times10^{-6}$, respectively (\SM). We have not found a case in which these small deviations matter
in practice. For small gaps or semicore states an optimal rational rule is the better choice. In the 3D window
implementation, the rule affects only the occupation function and not the cost (Section~\ref{sec:windows}).

\subsection{Accuracy}

\begin{figure*}[t]
\includegraphics[width=\textwidth]{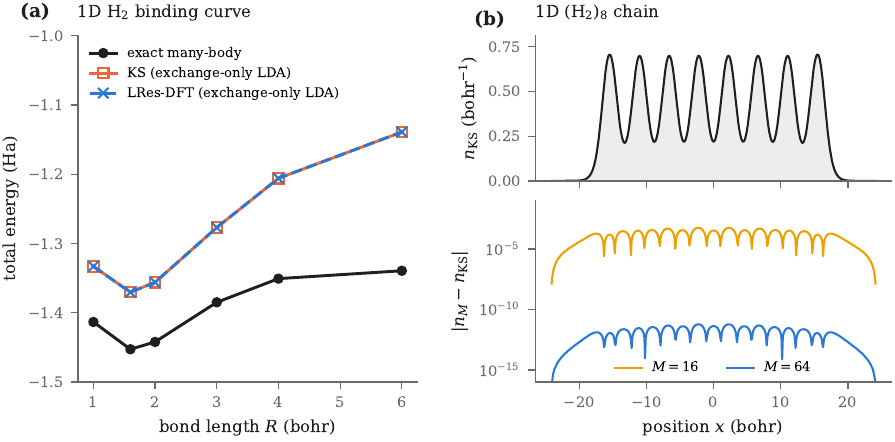}
\caption{\label{fig:acc}
(a)~Binding curve of 1D H$_2$: exact (black), KS (orange squares) and LRes-DFT (blue crosses), with the same exchange-only
LDA in KS and LRes-DFT. (b)~Self-consistent KS density of a 1D (H$_2$)$_8$ chain (top) and the LRes-DFT density error for $M=16$ and 64
(bottom, logarithmic scale).}
\end{figure*}

Table~\ref{tab:exact} and Figure~\ref{fig:acc}(a) compare LRes-DFT and KS, with the same exchange-only LDA, with the exact
solution of the interacting two-electron problem. LRes-DFT and KS agree to $5\times10^{-12}$~Ha or better, except for
stretched H$_2$ at $R=6$~bohr, where the gap nearly closes and the difference is $8\times10^{-5}$~Ha at $M=96$. Both
differ from the exact energies by 0.08--0.20~Ha, which is the error of the exchange functional. For LiH, Be and
dimerized (H$_2$)$_n$ chains, self-consistent LRes-DFT energies agree with KS to $6\times10^{-11}$~Ha per electron or better
(Table~\ref{tab:ks1d}). For a uniform H$_{20}$ chain with a gap of 0.05~Ha the agreement is $8\times10^{-8}$~Ha per
electron at $M=128$.

\begin{table}[t]
\caption{\label{tab:exact}
1D two-electron systems: total energies (Ha) from exact diagonalization, KS and LRes-DFT ($M=96$), with the same
exchange-only LDA in KS and LRes-DFT.}
\begin{ruledtabular}
\footnotesize\setlength{\tabcolsep}{2pt}
\begin{tabular}{lccccc}
System & Exact & KS & LRes & KS$-$ex. & $|$LRes$-$KS$|$ \\
\hline
He & -2.23855 & -2.15845 & -2.15845 & 0.0801 & $3.6\times10^{-15}$ \\
H$_2$, $R=1.0$ & -1.41320 & -1.33295 & -1.33295 & 0.0802 & $1.1\times10^{-14}$ \\
H$_2$, $R=1.6$ & -1.45258 & -1.37026 & -1.37026 & 0.0823 & $1.2\times10^{-14}$ \\
H$_2$, $R=2.0$ & -1.44200 & -1.35599 & -1.35599 & 0.0860 & $8.9\times10^{-16}$ \\
H$_2$, $R=3.0$ & -1.38483 & -1.27693 & -1.27693 & 0.1079 & $9.8\times10^{-15}$ \\
H$_2$, $R=4.0$ & -1.35067 & -1.20590 & -1.20590 & 0.1448 & $4.3\times10^{-12}$ \\
H$_2$, $R=6.0$ & -1.33930 & -1.13913 & -1.13905 & 0.2002 & $7.8\times10^{-5}$ \\

\end{tabular}
\end{ruledtabular}
\end{table}

\begin{table}[t]
\caption{\label{tab:ks1d}1D systems: self-consistent LRes-DFT ($M=128$) versus KS on the identical Hamiltonian.}
\begin{ruledtabular}
\footnotesize\setlength{\tabcolsep}{2pt}
\begin{tabular}{lcccc}
System & $N$ & $E_\mathrm{KS}$ (Ha) & $|\Delta E|/N$ (Ha) & $\|\Delta n\|_1/N$ \\
\hline
LiH & 4 & -4.836073 & $1.1\times10^{-12}$ & $5.0\times10^{-12}$ \\
Be & 4 & -6.692767 & $5.2\times10^{-11}$ & $2.5\times10^{-10}$ \\
(H$_2$)$_8$ chain & 16 & -10.983588 & $2.9\times10^{-15}$ & $5.1\times10^{-15}$ \\
(H$_2$)$_{32}$ chain & 64 & -43.944324 & $8.9\times10^{-16}$ & $4.9\times10^{-15}$ \\
H$_{20}$ chain (small gap) & 20 & -13.818405 & $7.7\times10^{-8}$ & $9.1\times10^{-7}$ \\

\end{tabular}
\end{ruledtabular}
\end{table}

\subsection{Forces}\label{sec:forces}

Table~\ref{tab:forces} tests Propositions~\ref{prop:var} and \ref{prop:rigid} for a 1D H$_2$ molecule. Finite differences
with steps of $10^{-2}$ and $10^{-3}$~bohr were extrapolated to zero step. At fixed $\mu$ and $E_\mathrm{lo}$, the derivative
of $G_M$ equals the Hellmann--Feynman force to $10^{-10}$~Ha/bohr for $M=8$, 16 and 64. At fixed $N$ with a rigidly moving
circle ($R=0.5$~Ha), the derivative of $A_M$ also equals the Hellmann--Feynman force to $10^{-10}$~Ha/bohr at every $M$, as
Proposition~\ref{prop:rigid} states. With $E_\mathrm{lo}$ fixed instead, the derivative of the fixed-$N$ energy,
Equation~\eqref{eq:energy}, differs from the Hellmann--Feynman force by $2.8\times10^{-3}$, $1.4\times10^{-5}$ and
$2\times10^{-12}$~Ha/bohr for $M=8$, 16 and 64. A rigidly moving circle should therefore be used when forces are needed at
small $M$. Forces in 3D, where the atom-centered basis adds basis-set (Pulay) force terms, were not computed.

\begin{table}[t]
\caption{\label{tab:forces}Force on a nucleus of 1D H$_2$: difference (Ha/bohr) between the Hellmann--Feynman force and the
extrapolated finite-difference derivative, at fixed $\mu$ (derivative of $G_M$), at fixed $N$ with a rigidly moving circle
(derivative of $A_M$), and at fixed $N$ with fixed $E_\mathrm{lo}$ (derivative of Equation~\eqref{eq:energy}).}
\begin{ruledtabular}
\footnotesize
\begin{tabular}{cccc}
$M$ & fixed $\mu$ & fixed $N$, rigid circle & fixed $N$, fixed $E_\mathrm{lo}$ \\
\hline
8 & $2.9\times10^{-11}$ & $6.1\times10^{-11}$ & $2.8\times10^{-3}$ \\
16 & $1.1\times10^{-10}$ & $2.5\times10^{-11}$ & $1.4\times10^{-5}$ \\
64 & $8.6\times10^{-11}$ & $8.7\times10^{-11}$ & $2.2\times10^{-12}$ \\

\end{tabular}
\end{ruledtabular}
\end{table}

\subsection{Cost}

LRes-DFT time grows as $N^{1.0}$--$N^{1.2}$ up to 8192 electrons [Figure~\ref{fig:cost}(a)]. In 1D, LRes-DFT with
Lemma~\ref{lem:recursion} is a pole expansion with selected inversion, which is known to scale linearly in one
dimension~\cite{PEXSI2013}. The comparison with KS eigensolvers therefore mainly confirms that the implementation reaches
this scaling. The KS solver (bisection with inverse iteration) grows as $N^{2.8}$ here, because the occupied states of a dimerized
chain form a narrow cluster. Our LRes-DFT timings include a secant search for $\mu$ that costs 25--47 density evaluations
when started without a previous value. The 1D window variant of LRes-DFT also scales linearly, with an error of
$(1.3$--$1.6)\times10^{-7}$ per electron from 64 to 4096 electrons.

\section{Results in three dimensions}\label{sec:res3d}

This section tests the window construction. We first measure the density error of a single window calculation at the
self-consistent KS potential, for crystals, disordered systems and compressed systems. We then compare the windows with
divide-and-conquer using Fermi--Dirac occupations, run complete SCF calculations, and measure the cost.

\subsection{Window error, disorder and compression}

Figure~\ref{fig:conv}(b) and Table~\ref{tab:compress} show the density error per electron as a function of the window
radius at the self-consistent KS potential. For the crystal at $a=6$~bohr the error is $3.0\times10^{-4}$, $1.0\times10^{-4}$,
$3.6\times10^{-6}$ and $3.2\times10^{-8}$ for $R_w=6.1$, 8.5, 12.1 and 13.5~bohr. The decrease is stepwise because the
molecules sit on a lattice, and the error is controlled by the nearest molecules outside the window.

In an amorphous arrangement at the same density, with windows centered on each molecule, the error at $R_w=12.1$~bohr is
$3.2\times10^{-5}$, nine times larger, and it decreases slowly ($1.3\times10^{-5}$ at 13.5~bohr). The self-consistent
energy agrees with KS to $2.7\times10^{-7}$~Ha per electron, with 20 SCF iterations for each method (convergence thresholds
$10^{-8}$ for KS and $10^{-7}$ for LRes-DFT).

Compression is more severe. At $a=5.5$, 5.0 and 4.5~bohr the gap falls to 0.23, 0.16 and 0.05~Ha. The error at
$R_w=12.1$~bohr then rises to $2.4\times10^{-5}$, $9.6\times10^{-5}$ and $9.5\times10^{-4}$, while the window grows to
400--970 basis functions. Increasing $M$ from 32 to 128 does not reduce the window error; it changes by at most a factor
of 1.5 and increases slightly with $M$. The window, not the occupation function, controls the accuracy in 3D.

\begin{table}[t]
\caption{\label{tab:compress}3D window error at the self-consistent KS potential ($M=64$) for crystals with lattice constant
$a$ (bohr) and an amorphous arrangement. $\Delta$: gap (Ha); $n_W$: basis functions per window; $\Delta E_\mathrm{b}$:
band-energy difference (Ha).}
\begin{ruledtabular}
\footnotesize\setlength{\tabcolsep}{2pt}
\begin{tabular}{lcccccc}
System & $N$ & $\Delta$ & $R_w$ & $n_W$ & $\|\Delta n\|_1/N$ & $|\Delta E_\mathrm{b}|/N$ \\
\hline
$a=6.0$ & 250 & 0.29 & 8.5 & 228 & $1.0\times10^{-4}$ & $2.0\times10^{-6}$ \\
$a=6.0$ & 250 & 0.29 & 12.1 & 396 & $3.6\times10^{-6}$ & $1.5\times10^{-7}$ \\
$a=5.5$ & 432 & 0.23 & 8.5 & 228 & $5.3\times10^{-4}$ & $4.7\times10^{-5}$ \\
$a=5.5$ & 432 & 0.23 & 12.1 & 396 & $2.4\times10^{-5}$ & $1.6\times10^{-7}$ \\
$a=5.0$ & 432 & 0.16 & 8.5 & 228 & $2.3\times10^{-3}$ & $2.2\times10^{-4}$ \\
$a=5.0$ & 432 & 0.16 & 12.1 & 684 & $9.6\times10^{-5}$ & $4.6\times10^{-6}$ \\
$a=4.5$ & 432 & 0.05 & 8.5 & 324 & $8.0\times10^{-3}$ & $5.5\times10^{-4}$ \\
$a=4.5$ & 432 & 0.05 & 12.1 & 972 & $9.5\times10^{-4}$ & $1.5\times10^{-5}$ \\
amorph. & 250 & 0.23 & 8.5 & 146 & $1.0\times10^{-3}$ & $6.7\times10^{-5}$ \\
amorph. & 250 & 0.23 & 12.1 & 414 & $3.2\times10^{-5}$ & $2.8\times10^{-7}$ \\

\end{tabular}
\end{ruledtabular}
\end{table}

The window error also depends on the basis (Table~\ref{tab:bases} and \SM). Compact and diffuse bases pull in opposite
directions: a more compact basis has a larger basis error but a smaller window error, and a more diffuse basis the reverse.
Diffuse functions make the overlap matrix nearly singular (smallest eigenvalue 0.004) and lengthen the range of the density
matrix in the basis. In all cases studied the window spectra contain no states inside the KS gap, and their lowest
eigenvalue lies above the global one, as Lemma~\ref{lem:interlace} requires (\SM).

\begin{table}[t]
\caption{\label{tab:bases}Basis sets for the 3D crystal: functions per atom, cutoff (bohr), basis error (Ha/electron)
relative to a converged grid calculation (16 electrons) and window error at $R_w=12.1$~bohr. Exponents (bohr$^{-2}$): compact, $s$ 1.0, 0.25 and $p$ 0.6; default, $s$ 1.2, 0.4, 0.15 and $p$ 0.5; diffuse, $s$ 1.0, 0.3, 0.12 and $p$ 0.5.}
\begin{ruledtabular}
\footnotesize\setlength{\tabcolsep}{3pt}
\begin{tabular}{lcccc}
Basis & Funcs./atom & Cutoff & Basis err. & Window err. \\
\hline
compact & 5 & 4.5 & 0.051 & $1.3\times10^{-7}$ \\
default & 6 & 4.5 & 0.019 & $3.6\times10^{-6}$ \\
diffuse & 6 & 5.5 & 0.005 & $3.8\times10^{-5}$ \\
\end{tabular}
\end{ruledtabular}
\end{table}

\subsection{Relation to divide-and-conquer}\label{sec:dc}

Because Equation~\eqref{eq:window} is a DC construction, we compared it with DC using Fermi--Dirac occupations
($k_BT=0.005$ and 0.01~Ha) in the same windows and at the same potential (Table~\ref{tab:dc}). With $k_BT=0.005$~Ha the
density errors agree with those obtained with $r_M$ to within 6\% at every window radius. With $k_BT=0.01$~Ha they agree to
within 10\%, except at the largest radius, where the smearing error of the Fermi--Dirac function exceeds the window error.
Both occupation functions are evaluated on the same window eigenpairs, so their cost is the same. In 3D the occupation
$r_M$ offers no accuracy advantage over Fermi--Dirac occupations. The speed-up over KS reported below is a property of
nearsighted DC methods in general.

\begin{table}[t]
\caption{\label{tab:dc}Density error per electron, 3D crystal (250 electrons), for the same windows with the occupation
$r_M$ ($M=64$) and with Fermi--Dirac occupations (DC).}
\begin{ruledtabular}
\footnotesize
\begin{tabular}{ccccc}
$R_w$ (bohr) & $n_W$ & $r_M$, $M=64$ & DC, $k_BT=0.005$ & DC, $k_BT=0.01$ \\
\hline
6.1 & 84 & $3.0\times10^{-4}$ & $3.0\times10^{-4}$ & $2.9\times10^{-4}$ \\
8.5 & 228 & $1.0\times10^{-4}$ & $1.0\times10^{-4}$ & $1.0\times10^{-4}$ \\
12.1 & 396 & $3.6\times10^{-6}$ & $3.4\times10^{-6}$ & $3.2\times10^{-6}$ \\
13.5 & 684 & $3.2\times10^{-8}$ & $3.0\times10^{-8}$ & $6.0\times10^{-8}$ \\

\end{tabular}
\end{ruledtabular}
\end{table}

\subsection{Self-consistent calculations}

\paragraph*{Accuracy.} Table~\ref{tab:scf3d} lists self-consistent results from 16 to 1024 electrons. For 16 and 54
electrons the window covers the whole periodic cell, so these rows test only the implementation. For 128 to 1024 electrons,
LRes-DFT energies agree with KS to between $1.6\times10^{-7}$ and $2.4\times10^{-7}$~Ha per electron. The density difference, about
$4.5\times10^{-6}$ per electron, does not grow with system size, and both methods need the same number of SCF iterations
(11--13). At 432 electrons the self-consistent energy error is $3.7\times10^{-6}$, $1.6\times10^{-7}$ and $8.1\times10^{-9}$~Ha per electron for $R_w=8.5$, 12.1 and
13.5~bohr, so the window radius controls the self-consistent error as it does the non-self-consistent one.

\paragraph*{Scaling and break-even.} Table~\ref{tab:rigorous} gives the wall time of complete SCF calculations. The time
scales as $N^{2.92\pm0.02}$ for dense KS and $N^{1.15\pm0.06}$ for LRes-DFT. The fits use the four largest KS sizes and the five largest LRes-DFT sizes,
including one LRes-DFT run at 2000 electrons; the quoted error is the least-squares error of the slope only. Dense KS is
faster below about 500 electrons, by a factor of 3 to 6 at 54--250 electrons. LRes-DFT is 2.0
times faster at 686 electrons and 3.5 times faster at 1024. Extrapolating the KS fit to 2000
electrons gives about $3\times10^{3}$~s, against 272~s measured for LRes-DFT. We did not run KS at that size, so this
extrapolation should not be quoted as a result.

\paragraph*{Reliability of the timings.} The three repeats of each size differ by less than 7\%. Single timings taken earlier on a shared machine fluctuated by up to 1.5$\times$ (44 and
60~s for LRes-DFT at 432 electrons) and cannot be compared with these. The reference is single-threaded dense LAPACK
(\texttt{dsygvx}). We did not time ELPA, a sparse solver or any other linear-scaling code, and a faster dense solver or a
better-preconditioned iterative solver would move the break-even point to larger $N$. With eight threads, the KS step at
432 electrons went from 2.9 to 2.3~s and the LRes-DFT step from 2.7 to 1.0~s; these threaded timings come from a shared
machine.

\paragraph*{Accuracy against cost.} The radius used for the headline numbers, $R_w=12.1$~bohr, leaves an energy error of
about $2\times10^{-7}$~Ha per electron. A smaller error needs a larger window and more time: at 432 electrons, an energy error
of $10^{-8}$~Ha per electron takes 159~s at $R_w=13.5$~bohr against 44~s at 12.1~bohr and 39~s for KS (a separate run on a loaded machine), i.e.\ it is slower than dense KS.

\paragraph*{Scope of the timings.} All timings are for a compact Gaussian basis (six functions per electron) with local
pseudopotentials, and $M=64$ poles. Fewer poles would reduce the cost in proportion and increase the error
(Section~\ref{sec:rules}). We did not time nonlocal pseudopotentials, larger basis sets or diffuse functions, which enlarge
the windows. The accompanying code also treats small molecules with GTH pseudopotentials (up to 10 atoms, 8 electrons). There
the window covers the whole molecule and KS and LRes-DFT take the same time, about 7~s for H$_2$O at a grid spacing of
0.2~bohr, which is dominated by grid operations common to both. No speed-up is expected for such molecules.

\paragraph*{Iterative reference.} The warm-started iterative KS solver was slower than dense diagonalization in this basis
(128 electrons: 3~s versus 2~s; 432 electrons: 44~s versus 34~s; at 1024 electrons a warm LOBPCG step took 46~s (up to 63 iterations); \SM). The occupied states make up 1/12 of the basis, and the Rayleigh--Ritz step of LOBPCG on a
block of about $3N/2$ vectors costs nearly as much as a dense diagonalization. Other iterative solvers or optimized dense
libraries such as ELPA~\cite{ELPA2014} may perform differently; we have not tested them.

\begin{table}[t]
\caption{\label{tab:rigorous}Wall time (s, median and range of three runs) of a complete SCF calculation, single thread,
idle machine, 11 SCF iterations for both methods; the last column is the ratio of the medians. The 2000-electron LRes-DFT
value is a single run; dense KS was not run at that size.}
\begin{ruledtabular}
\footnotesize
\begin{tabular}{rccc}
$N$ & KS & LRes-DFT & KS/LRes \\
\hline
54 & 0.5 (0.5--0.6) & 3.2 (3.2--3.3) & 0.17 \\
128 & 1.9 (1.9--2.0) & 8.6 (8.4--8.9) & 0.22 \\
250 & 7.2 (7.1--7.4) & 25.0 (24.7--25.2) & 0.29 \\
432 & 34.6 (33.9--35.2) & 46.0 (44.6--46.5) & 0.75 \\
686 & 132.0 (131.4--136.5) & 65.8 (65.4--66.4) & 2.01 \\
1024 & 442.9 (441.2--445.1) & 125.4 (125.0--126.0) & 3.53 \\
2000 & -- & 272.1 (1 run) & -- \\
\end{tabular}
\end{ruledtabular}
\end{table}

\begin{table*}[t]
\caption{\label{tab:scf3d}Self-consistent 3D results: LRes-DFT ($M=64$, $R_w=12.1$~bohr) versus dense KS in the identical
basis. Wall times are for the complete SCF calculation (single thread), from one run each on a shared machine. Use
Table~\ref{tab:rigorous}, which has repeated timings, for speed-ups.}
\begin{ruledtabular}
\footnotesize
\begin{tabular}{cccccccc}
$N$ & $E_\mathrm{KS}$ (Ha) & $|\Delta E|/N$ (Ha) & $\|\Delta n\|_1/N$ & SCF it.\ KS/LRes & KS wall (s) & LRes wall (s) & KS/LRes \\
\hline
16 & -4.634688 & $6.5\times10^{-10}$ & $9.7\times10^{-9}$ & 13/17 & 0.2 & 0.4 & 0.51 \\
54 & -15.746533 & $2.1\times10^{-13}$ & $8.7\times10^{-9}$ & 13/16 & 0.7 & 5.6 & 0.13 \\
128 & -37.357441 & $2.4\times10^{-7}$ & $4.8\times10^{-6}$ & 13/13 & 2.5 & 14.8 & 0.17 \\
432 & -126.019962 & $1.6\times10^{-7}$ & $4.6\times10^{-6}$ & 11/11 & 39 & 44 & 0.89 \\
686 & -200.257102 & $1.9\times10^{-7}$ & $4.4\times10^{-6}$ & 11/11 & 136 & 62 & 2.20 \\

\end{tabular}
\end{ruledtabular}
\end{table*}

\subsection{Cost of a density build, memory and threads}

\begin{figure*}[t]
\includegraphics[width=\textwidth]{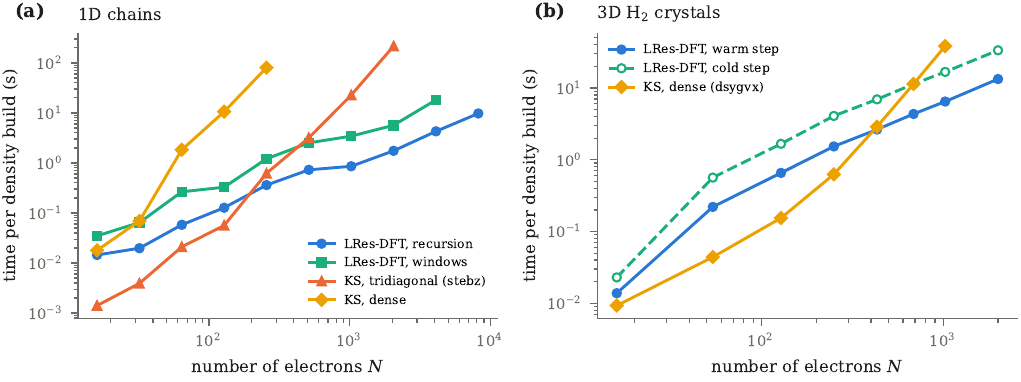}
\caption{\label{fig:cost}
Time for one density build (single thread). (a)~1D dimerized chains: LRes-DFT with the exact recursion and with local
windows, and KS with a tridiagonal eigensolver and with dense diagonalization. (b)~3D H$_2$ crystals in the identical basis:
LRes-DFT warm and cold steps and dense KS (\texttt{dsygvx}).}
\end{figure*}

Figure~\ref{fig:cost}(b) and Table~\ref{tab:cost3d} give the cost of one density build. The fitted exponents are
$N^{2.92}$ for KS (250--1024 electrons), $N^{1.05}$ for the LR-DFT warm step and $N^{1.02}$ for the cold step (432--2000 electrons). The KS time is dominated by the reduction of the dense matrices to tridiagonal form, which
costs $\mathcal O(N_b^3)$ for all three LAPACK paths tested. Building the density takes at most 8\% of the KS time (\SM).

The LRes-DFT prefactor is large, because every molecule lies in about 33 overlapping windows of about 400 basis functions
each. Within a window, the eigendecomposition is much cheaper than the equivalent resolvent solves: for 396 basis functions it
took 0.024~s, against 0.46--0.59~s for 32 complex solves, and the ratio was similar at 228 and 684 basis functions. There are three break-even points. For a single density build the crossover is at about 411 electrons for the warm step and about 685 electrons for the cold step; at 1024 electrons LR-DFT is 5.9 (warm) and 2.3 (cold) times faster. A complete SCF calculation consists of one cold step and several warm
steps, so its break-even lies between them, near 500 electrons (Table~\ref{tab:rigorous}).
This last figure is the one we quote as the headline.

The additional peak memory of the solver step grows as $N^{2.1}$ for KS and $N^{1.3}$ for LR-DFT; at 1024 electrons the solver step needs 676~MB (KS) and 251~MB (LR-DFT) in addition to the shared basis and grid data (\SM). For 250 electrons and below the two
methods use comparable memory. The windows are independent, and with OpenMP over windows the LRes-DFT step at 432 electrons
went from 2.66 to 0.97~s on 8 threads (2.7$\times$), while the threaded LAPACK of the KS step went from 2.90~s to 2.26~s at best (\SM). These thread timings are rough, because the machine was shared and the threaded
OpenBLAS is not the best-performing dense solver library.

\begin{table}[t]
\caption{\label{tab:cost3d}3D cost of one density build (s, single thread, best of two or three) and the LRes-DFT--KS density
difference per electron at the initial-guess potential. $N_b$: basis functions.}
\begin{ruledtabular}
\footnotesize\setlength{\tabcolsep}{3pt}
\begin{tabular}{rrrrrr}
$N$ & $N_b$ & KS & LRes warm & LRes cold & $\|\Delta n\|_1/N$ \\
\hline
16 & 96 & 0.01 & 0.01 & 0.02 & $3.3\times10^{-15}$ \\
54 & 324 & 0.04 & 0.22 & 0.56 & $1.7\times10^{-15}$ \\
128 & 768 & 0.15 & 0.66 & 1.67 & $8.0\times10^{-7}$ \\
250 & 1500 & 0.63 & 1.53 & 4.07 & $6.3\times10^{-7}$ \\
432 & 2592 & 2.88 & 2.64 & 6.94 & $7.2\times10^{-7}$ \\
686 & 4116 & 11.35 & 4.34 & 11.33 & $7.9\times10^{-7}$ \\
1024 & 6144 & 38.20 & 6.46 & 16.73 & $7.8\times10^{-7}$ \\
2000 & 12000 & -- & 13.30 & 33.41 & -- \\

\end{tabular}
\end{ruledtabular}
\end{table}

\section{Relation to other methods}\label{sec:comparison}

Table~\ref{tab:methods} places LRes-DFT among existing approaches. Only the comparisons with KS (dense and LOBPCG), with DC
using Fermi--Dirac occupations, and with the Gauss--Legendre and Zolotarev rules were made in this work.

The formulation is closest to the Green's-function methods. KKR-based methods such as LSMS and LSGF~\cite{LSMS1995,LSGF1996,Zeller2008}
also compute site-diagonal Green's functions on a complex contour in local zones, and they obtain the band energy from a
contour integral. They use a multiple-scattering representation, whereas here the resolvent is taken in a grid or localized basis. Pole
expansions with selected inversion~\cite{LinLuYingE2009,PEXSI2013} evaluate the same contour integral with optimal poles and
no locality truncation. Our 1D implementation is of this type with a non-optimal rule, and optimal rules need fewer poles
(Section~\ref{sec:rules}). In 3D our implementation is a DC method~\cite{Yang1991}, equivalent in accuracy and cost to DC with
Fermi--Dirac occupations (Section~\ref{sec:dc}). Fragment methods such as LS3DF~\cite{LS3DF2008} use a patching scheme that
cancels boundary errors, which we have not tested. Compared with orbital-free DFT, LRes-DFT keeps the exact KS kinetic
energy, at a much larger cost per electron.

\begin{table*}[t]
\caption{\label{tab:methods}
LRes-DFT and related approaches. ``Extra error'' is the error beyond that of the xc functional and basis. Entries other than
KS, DC and LRes-DFT summarize the cited literature and were not measured here.}
\begin{ruledtabular}
\footnotesize
\begin{tabular}{p{4.0cm}p{2.7cm}p{2.0cm}p{2.3cm}p{5.6cm}}
Method & Variable & Kinetic energy & Cost (3D, gapped) & Extra error and its control \\
\hline
KS, dense diagonalization & $N/2$ orbitals & exact & $\mathcal O(N^3)$ & none (measured here: $N^{2.9}$) \\
Orbital-free DFT~\cite{WangTeter1992,WGC1999,HuangCarter2010} & $n(\bm r)$ & approximate & $\mathcal O(N\ln N)$ & kinetic functional \\
Density matrix / localized orbitals~\cite{LNV1993,ONETEP2005,MGC1993,BigDFTLS2014} & sparse density matrix or orbitals & exact & $\mathcal O(N)$ & localization radius \\
Divide-and-conquer, fragments~\cite{Yang1991,LS3DF2008} & local densities & exact & $\mathcal O(N)$ & buffer radius (DC measured here) \\
Chebyshev / Fermi operator~\cite{Goedecker1999,WeisseKPM2006} & polynomial of $\hat h$ & exact & $\mathcal O(N)$ & degree $\propto$ spectral width \\
Pole expansion + selected inversion~\cite{LinLuYingE2009,PEXSI2013,Moussa2016} & resolvents & exact & $\mathcal O(N^2)$ & pole number $\propto\ln(W/\Delta)$ \\
KKR-based (LSMS, LSGF)~\cite{LSMS1995,LSGF1996,Zeller2008} & site Green's functions & exact & $\mathcal O(N)$ & local-zone size \\
Stochastic DFT~\cite{BaerSDFT2013} & random traces & exact & $\lesssim\mathcal O(N)$ & statistical noise \\
LRes-DFT (this work) & local resolvent $g$ & exact & $\mathcal O(N)$ & $\rho^M$ (rule) and window radius \\
\end{tabular}
\end{ruledtabular}
\end{table*}

\section{Limitations}\label{sec:limits}

\begin{description}
\item[Small gaps and metals.] The number of poles of the half-step rule grows as $W_\mathrm{occ}/\Delta$, and the window
error grows as the gap closes. At a gap of 0.05~Ha (compressed H$_2$) the 3D density error is $9.5\times10^{-4}$ per electron
at $R_w=12.1$~bohr. Metals, and with them the alloy problems that motivate this work, were not studied. They would need
finite electronic temperature and an optimal pole rule.
\item[Disorder and basis set.] The window error is nine times larger for an amorphous arrangement than for the crystal,
and it grows with the diffuseness of the basis.
\item[Scope.] Only local pseudopotentials and the LDA were used. Nonlocal operators need off-diagonal density-matrix
elements (Remark~\ref{rem:potential}). The 3D tests use only H$_2$ molecular solids. The basis has its own error of
0.019~Ha per electron.
\item[Forces.] Forces were verified in 1D only. 3D forces, which need the basis-set (Pulay) terms, were not computed.
\item[Prefactor and references.] In 3D, LRes-DFT is slower than dense KS below the break-even size, and its gain above it is
modest (Table~\ref{tab:rigorous}). The gain relative to a well-tuned dense or iterative solver, and for larger or
nonlocal-pseudopotential bases, is unknown. We did not compare with ELPA~\cite{ELPA2014}, ELSI~\cite{ELSI2018} or other
linear-scaling codes, and parallel efficiency was not measured on a dedicated machine.
\end{description}

\section{Conclusions}

We have written the Kohn--Sham density in terms of the diagonal of the one-electron resolvent and analyzed a half-step
trapezoid rule for it. The density and band energy are linear in the local resolvent, the occupations lie in $[0,1]$ for
every $M$, and a circle that moves rigidly with the chemical potential gives exact fixed-$N$ forces. In one dimension the
error obeys an a priori bound that depends on the grid only logarithmically. Gauss--Legendre and Zolotarev rules need fewer
nodes, and the cost of the half-step rule grows with the ratio of occupied band width to gap.

In three dimensions the method is a divide-and-conquer scheme with the same accuracy as divide-and-conquer with
Fermi--Dirac occupations. For gapped H$_2$ molecular solids it reproduces self-consistent Kohn--Sham energies to
$2.5\times10^{-7}$~Ha per electron up to 1024 electrons. The cost grows linearly with system size, but the prefactor is
large: on one core the method breaks even with dense diagonalization near 500 electrons and is
3.5 times faster at 1024 electrons. The window error grows with disorder, compression and basis
diffuseness, and it is the main limitation in three dimensions. Extending the approach to metals will require finite
electronic temperature and an optimal pole rule.

\begin{acknowledgments}
This work received no external funding. The AI assistant Claude Sonnet 5.5 (Anthropic) was also used to search the
literature and to draft and edit the text; the uses in the code and calculations are described in
Sec.~\ref{sec:details}. The author directed this work, checked the results and the text, and takes full responsibility
for the content. The AI tool is not an author.

\emph{Author contributions} (CRediT). A.L.: conceptualization, methodology, software, validation, formal analysis,
investigation, data curation, visualization, writing -- original draft, writing -- review and editing.

\emph{Competing interests.} The author declares no competing interests.

\emph{Data availability.} The code, the computer-algebra scripts, the tests, the benchmark scripts and all data used in the
figures and tables are available in Ref.~\cite{LResDFTcode}.
\end{acknowledgments}

\appendix

\section{Proofs}\label{app:proofs}

Notation is that of Section~\ref{sec:theory}: $u_i=(\varepsilon_i-c)/R$, $r_M(u)=1/(1+u^M)$, $\zeta_j=\ee^{\ii\pi(2j+1)/M}$
($j=0,\dots,M-1$), $z_j=c+R\zeta_j$ and $w_j=R\zeta_j/M$. Two elementary facts are used repeatedly.
\begin{enumerate}
\item[(F1)] $\zeta_j^M=\ee^{\ii\pi(2j+1)}=-1$, and the $M$ numbers $\zeta_j$ are distinct, so they are \emph{all} the roots of
$\zeta^M+1$, and each is simple.
\item[(F2)] $\sum_j\zeta_j=0$, because $\zeta_j=\zeta_0\,\omega^{j}$ with $\omega=\ee^{2\pi\ii/M}\neq1$, and
$\sum_j\omega^j=(1-\omega^M)/(1-\omega)=0$. Hence $\sum_jw_j=0$.
\end{enumerate}
Also, $\zeta_{M-1-j}=\overline{\zeta_j}$, so the nodes and weights come in complex-conjugate pairs.

\subsection{Closed form and bounds (Lemmas~\ref{lem:closed} and~\ref{lem:bounds})}

\begin{proof}[Proof of Lemma~\ref{lem:closed}]
\emph{Step 1: residues.} Write $f(u)=p(u)/q(u)$ with $p=1$ and $q=1+u^M$. By (F1) the zeros of $q$ are simple, so the
residue at $\zeta_j$ is $p/q'$:
\begin{align}
\operatorname{Res}_{u=\zeta_j}f
&=\frac{1}{M\zeta_j^{M-1}}
 =\frac{\zeta_j}{M\,\zeta_j^{M}}\nonumber\\
&=-\frac{\zeta_j}{M}.
\end{align}

\emph{Step 2: partial fractions.} Let $s(u)=\sum_j\operatorname{Res}_{\zeta_j}f/(u-\zeta_j)$. The difference $f-s$ has no
poles, so it is entire; both $f$ and $s$ tend to $0$ as $|u|\to\infty$, so $f-s$ is bounded and tends to $0$. By
Liouville's theorem, $f-s\equiv0$:
\begin{equation}
f(u)=-\frac1M\sum_j\frac{\zeta_j}{u-\zeta_j}=\frac1M\sum_j\frac{\zeta_j}{\zeta_j-u}.
\end{equation}

\emph{Step 3: change of variable.} Put $u=(\varepsilon-c)/R$. Multiplying numerator and denominator by $R$ and using
$R\zeta_j=z_j-c$,
\begin{align}
\frac{1}{M}\frac{\zeta_j}{\zeta_j-u}
&=\frac{R\zeta_j}{M}\,\frac{1}{R\zeta_j-(\varepsilon-c)}\nonumber\\
&=\frac{w_j}{z_j-\varepsilon}.
\end{align}
Summing over $j$ gives the scalar identity
\begin{equation}
r_M(u)=\sum_j\frac{w_j}{z_j-\varepsilon}.
\label{eq:scalarid}
\end{equation}

\emph{Step 4: matrix form.} Let $\hat h=\sum_i\varepsilon_i|\phi_i\rangle\langle\phi_i|$. Then
$(z_j-\hat h)^{-1}=\sum_i|\phi_i\rangle\langle\phi_i|/(z_j-\varepsilon_i)$, and by Equation~\eqref{eq:scalarid}
\begin{equation}
\sum_jw_j(z_j-\hat h)^{-1}=\sum_ir_M(u_i)\,|\phi_i\rangle\langle\phi_i|=r_M\Big(\frac{\hat h-c}{R}\Big),
\end{equation}
which is Equation~\eqref{eq:PM}.

\emph{Step 5: non-orthogonal basis.} Since $S>0$, the matrix $S^{-1}H=S^{-1/2}\big(S^{-1/2}HS^{-1/2}\big)S^{1/2}$ is similar
to a real symmetric matrix, so $S^{-1}H=X\Lambda X^{-1}$ with real diagonal $\Lambda$. Then
$(zS-H)^{-1}=(z-S^{-1}H)^{-1}S^{-1}=X(z-\Lambda)^{-1}X^{-1}S^{-1}$, and applying Equation~\eqref{eq:scalarid} to each diagonal
entry of $\Lambda$ gives $\sum_jw_j(z_jS-H)^{-1}=X\,r_M(\Lambda)X^{-1}S^{-1}=r_M(S^{-1}H)S^{-1}$.
\end{proof}

\begin{proof}[Proof of Lemma~\ref{lem:bounds}]
\emph{Step 1: range.} For real $u$ and even $M$, $u^M=(u^2)^{M/2}=|u|^M\ge0$. Therefore $1+u^M\ge1$ and $0<r_M\le1$.
The eigenvalues of $\hat P_M$ are the numbers $r_M(u_i)\in(0,1]$, so $\hat P_M$ is Hermitian with $0\preceq\hat P_M\preceq1$.
Together with the factor $2$ for spin, this is the definition of an ensemble-representable density matrix.

\emph{Step 2: states inside the circle.} If $|u|\le\rho<1$, then
\begin{equation}
1-r_M=\frac{|u|^M}{1+|u|^M}\le|u|^M\le\rho^M .
\end{equation}

\emph{Step 3: states outside the circle.} If $|u|\ge1/\rho>1$, then
\begin{equation}
r_M=\frac{1}{1+|u|^M}\le|u|^{-M}\le\rho^M .
\end{equation}

\emph{Example.} For $M=16$ and $u=0.8$: $u^M=0.0282$ and $r_M=0.973$; for $u=1.25$: $u^M=35.5$ and $r_M=0.027$.
For $M=64$ these become $1-6.3\times10^{-7}$ and $6.3\times10^{-7}$.
\end{proof}

\subsection{Density and band energy (Proposition~\ref{prop:rep}, Remark~\ref{rem:potential})}

\begin{proof}[Proof of Proposition~\ref{prop:rep}]
\emph{Step 1: residues of $g$.} By Equation~\eqref{eq:g}, $g(\bm r,z)=\sum_i|\phi_i(\bm r)|^2/(z-\varepsilon_i)$ has a simple pole at
each $\varepsilon_i$ with residue $|\phi_i(\bm r)|^2$. The function $z\,g$ has the residue
$\varepsilon_i|\phi_i(\bm r)|^2$ there, because $z=\varepsilon_i$ at the pole.

\emph{Step 2: residue theorem.} For a counterclockwise contour $C$ that encloses the occupied levels and no others,
\begin{align}
\frac{1}{2\pi\ii}\oint_Cg\,dz&=\sum_{i\,\mathrm{occ}}|\phi_i(\bm r)|^2,\\
\frac{1}{2\pi\ii}\oint_Cz\,g\,dz&=\sum_{i\,\mathrm{occ}}\varepsilon_i|\phi_i(\bm r)|^2 .
\end{align}
Multiplying the first by $2$ gives (a).

\emph{Step 3: sum over grid points.} Because $\sum_{\bm r}|\phi_i(\bm r)|^2=1$ for each normalized $\phi_i$, summing the
second line over $\bm r$ and multiplying by $2$ gives (b).

\emph{Step 4: the discretized rule.} Replacing $\frac{1}{2\pi\ii}\oint_C(\cdot)\,dz$ by $\sum_jw_j(\cdot)|_{z_j}$ and using
Lemma~\ref{lem:closed}, the density is
$n_M\Delta V=2\sum_jw_jg(\bm r,z_j)=2\sum_ir_M(u_i)|\phi_i(\bm r)|^2$. For the band energy, write
$\hat h=z_j-(z_j-\hat h)$ inside the trace:
\begin{align}
\Tr[\hat h\hat P_M]
&=\sum_jw_j\Tr\big[\hat h\,(z_j-\hat h)^{-1}\big]\nonumber\\
&=\sum_jw_j\Tr\big[z_j(z_j-\hat h)^{-1}-1\big]\nonumber\\
&=\sum_jw_jz_j\Tr(z_j-\hat h)^{-1}-N_g\sum_jw_j .
\end{align}
By (F2) the last term vanishes. Since $\Tr(z-\hat h)^{-1}=\sum_{\bm r}g(\bm r,z)$, the band energy
$2\Tr[\hat h\hat P_M]=2\sum_i\varepsilon_ir_M(u_i)$ is again linear in $g$.
\end{proof}

\begin{proof}[Proof of Remark~\ref{rem:potential}]
\emph{Step 1: Neumann series.} For $|z|>\|\hat h\|$,
$(z-\hat h)^{-1}=z^{-1}(1-\hat h/z)^{-1}=\sum_{k\ge0}\hat h^k/z^{k+1}$, convergent in operator norm.

\emph{Step 2: expansion of $z\,g$.} Taking the $\langle\bm r|\cdot|\bm r\rangle$ element and multiplying by $z$,
\begin{equation}
z\,g(\bm r,z)=1+\frac{h_{\bm r\bm r}}{z}+\frac{(\hat h^2)_{\bm r\bm r}}{z^2}+\cdots .
\end{equation}

\emph{Step 3: contour integral.} On $|z|=R'$ with $R'>\|\hat h\|$, termwise integration is allowed. Only the $1/z$ term
survives, since $\frac{1}{2\pi\ii}\oint z^{-k}dz=\delta_{k1}$, and it gives $h_{\bm r\bm r}$. Finally
$h_{\bm r\bm r}=t_{\bm r\bm r}+v(\bm r)$ for a local potential.
\end{proof}

\subsection{Stationarity and forces (Propositions~\ref{prop:var} and~\ref{prop:rigid})}

\begin{proof}[Proof of Proposition~\ref{prop:var} and of the stationarity of $G_M$]
\emph{Step 1: properties of $W_M$.} Differentiating term by term,
\begin{equation}
W_M'(\varepsilon)=2\sum_j\frac{w_j}{z_j-\varepsilon}=2r_M(u),
\end{equation}
by Equation~\eqref{eq:scalarid}. No node is real, so each logarithm is analytic along the real axis. The conjugate pairing of
nodes and weights makes the sum real.

\emph{Step 2: derivative of a spectral sum.} Let $\Omega_M=\sum_iW_M(\varepsilon_i)$. First-order perturbation theory gives
$\partial\varepsilon_i/\partial v(\bm r)=\langle\phi_i|\bm r\rangle\langle\bm r|\phi_i\rangle=|\phi_i(\bm r)|^2$, because
$\partial\hat h/\partial v(\bm r)=|\bm r\rangle\langle\bm r|$. Hence
\begin{align}
\frac{\partial\Omega_M}{\partial v(\bm r)}
&=\sum_iW_M'(\varepsilon_i)\,|\phi_i(\bm r)|^2\nonumber\\
&=2\sum_ir_M(u_i)|\phi_i(\bm r)|^2\nonumber\\
&=n_M(\bm r)\Delta V .
\label{eq:dOmega}
\end{align}
For a single pole, the same argument is Jacobi's formula
$\partial\ln\det(z-\hat h)/\partial v(\bm r)=-g(\bm r,z)$.

\emph{Step 3: derivative of $G_M$.} Define $\chi_{\bm r\bm r'}=\partial n_M(\bm r)/\partial v(\bm r')$. In
Equation~\eqref{eq:GM}, $E_\mathrm H$ and $E_\mathrm{xc}$ depend on $v$ only through $n_M$, with
$\partial(E_\mathrm H+E_\mathrm{xc})/\partial n(\bm r)=v_\mathrm{Hxc}(\bm r)\Delta V$. The product rule on the middle term gives
\begin{align}
\frac{\partial G_M}{\partial v(\bm r')}
={}&n_M(\bm r')\Delta V-n_M(\bm r')\Delta V\nonumber\\
&-\sum_{\bm r}(v-v_\mathrm{ext})(\bm r)\,\chi_{\bm r\bm r'}\Delta V\nonumber\\
&+\sum_{\bm r}v_\mathrm{Hxc}[n_M](\bm r)\,\chi_{\bm r\bm r'}\Delta V .
\end{align}
The first line uses Equation~\eqref{eq:dOmega} and cancels. Collecting the rest,
\begin{equation}
\frac{\partial G_M}{\partial v(\bm r')}
=\sum_{\bm r}\chi_{\bm r\bm r'}\big[v_\mathrm{ext}+v_\mathrm{Hxc}[n_M]-v\big](\bm r)\,\Delta V .
\end{equation}
The bracket vanishes at self-consistency, $v=v_\mathrm{ext}+v_\mathrm{Hxc}[n_M]$, so $G_M$ is stationary. No assumption on
$\chi$ is needed.

\emph{Step 4: force.} Let $X$ be a nuclear coordinate. At fixed $v$, $\Omega_M$ and $n_M$ do not depend on $X$, and $E_\mathrm H$ and
$E_\mathrm{xc}$ depend on $X$ only through $n_M$. So $X$ enters $G_M$ only through $v_\mathrm{ext}$ in the middle term and
through $E_{nn}$. By the chain rule and Step~3,
\begin{align}
\frac{dG_M}{dX}
&=\frac{\partial G_M}{\partial X}\Big|_v+\sum_{\bm r'}\underbrace{\frac{\partial G_M}{\partial v(\bm r')}}_{=\,0}\frac{dv(\bm r')}{dX}\nonumber\\
&=\sum_{\bm r}n_M\,\frac{\partial v_\mathrm{ext}}{\partial X}\,\Delta V+\frac{\partial E_{nn}}{\partial X},
\end{align}
which is the Hellmann--Feynman expression.
\end{proof}

\begin{proof}[Proof of Proposition~\ref{prop:rigid}]
\emph{Step 1: $\mu$-dependence of the rule.} With $R$ fixed and $c=\mu-R$, the nodes are $z_j=\mu-R+R\zeta_j$ and the
weights $w_j=R\zeta_j/M$ do not depend on $\mu$. In Equation~\eqref{eq:scalarid} only $u=(\varepsilon-\mu)/R+1$ depends on $\mu$, so
$W_M$ is a function of $\varepsilon-\mu$ and
\begin{equation}
\frac{\partial W_M}{\partial\mu}=-W_M'=-2r_M .
\end{equation}

\emph{Step 2: electron number.} Summing over the eigenvalues,
\begin{align}
\frac{\partial\Omega_M}{\partial\mu}&=-\sum_i2r_M(u_i)=-N_M,\\
\frac{\partial A_M}{\partial\mu}&=-N_M+N,
\end{align}
which vanishes when $N_M=N$.

\emph{Step 3: stationarity in $v$.} At fixed $\mu$ the circle is fixed, so Step~3 of the previous proof applies unchanged, and
$\mu N$ does not depend on $v$.

\emph{Step 4: force.} $A_M$ is stationary with respect to $v$ and $\mu$ at the self-consistent solution, so its total
derivative with respect to $X$ equals its explicit derivative, which is the Hellmann--Feynman expression of the previous
proof. The force is $-dA_M/dX$, exactly at every $M$.
\end{proof}

\subsection{Convergence (Theorem~\ref{thm:conv})}

\begin{proof}[Proof of Theorem~\ref{thm:conv}]
\emph{Idea.} The density error is bounded by the sum of the occupation errors of all eigenstates. Occupied states lie inside the
circle and each contributes at most $\rho^M$. Empty states close to the circle also contribute at most $\rho^M$, and empty
states far from it at most $2^{-M}$.

\emph{Step 1: the circle in terms of the gap.} From $E_\mathrm{lo}=c-R$ and $\mu=c+R$ we get
$2R=\mu-E_\mathrm{lo}=\Delta_-+(\varepsilon_\mathrm{HOMO}-E_\mathrm{min})+\delta$. The scaled energies of the three landmark
states are
\begin{align}
u_\mathrm{min}&=\frac{E_\mathrm{min}-c}{R}=-1+\frac{\delta}{R},\nonumber\\
u_\mathrm{HOMO}&=\frac{\varepsilon_\mathrm{HOMO}-\mu}{R}+1=1-\frac{\Delta_-}{R},\nonumber\\
u_\mathrm{LUMO}&=\frac{\varepsilon_\mathrm{LUMO}-\mu}{R}+1=1+\frac{\Delta_+}{R}.
\end{align}

\emph{Step 2: reduction to eigenvalues.} Let $\hat P$ be the exact projector on the occupied states and
$A=\hat P_M-\hat P$. Then $(n_M-n)\Delta V=2\,\mathrm{diag}\,A$. $A$ has the eigenvectors $\phi_i$ of $\hat h$ and
eigenvalues $a_i=r_M(u_i)-\theta_i$, with $\theta_i=1$ for occupied and $0$ for empty states. By the triangle inequality
and normalization,
\begin{align}
\sum_{\bm r}|A_{\bm r\bm r}|
&=\sum_{\bm r}\Big|\sum_ia_i|\phi_i(\bm r)|^2\Big|\nonumber\\
&\le\sum_i|a_i|\sum_{\bm r}|\phi_i(\bm r)|^2\nonumber\\
&=\sum_i|a_i| .
\end{align}

\emph{Step 3: occupied states.} The map $\varepsilon\mapsto u$ is increasing, so occupied states lie in
$[u_\mathrm{min},u_\mathrm{HOMO}]$ and $|u_i|\le\max(|1-\delta/R|,1-\Delta_-/R)\le\rho$. By Lemma~\ref{lem:bounds},
$|a_i|=1-r_M\le\rho^M$. There are $N_\mathrm{occ}=N/2$ such states.

\emph{Step 4: empty states.} These satisfy $u_i\ge u_\mathrm{LUMO}=1+\Delta_+/R$, hence $|u_i|^{-1}\le1/(1+\Delta_+/R)\le\rho$. In
the first group, $1<|u_i|<2$ ($N_\mathrm{near}$ states), Lemma~\ref{lem:bounds} gives $|a_i|=r_M\le\rho^M$. In the second,
$|u_i|\ge2$ (at most $N_g$ states), $|a_i|=r_M\le|u_i|^{-M}\le2^{-M}$.

\emph{Step 5: sum.} Adding the three contributions and multiplying by $2$,
\begin{equation}
\sum_{\bm r}|n_M-n|\Delta V\le2\,(N_\mathrm{occ}+N_\mathrm{near})\,\rho^M+2N_g\,2^{-M}.
\end{equation}
\end{proof}

\subsection{Recursion and window spectra (Lemmas~\ref{lem:recursion} and~\ref{lem:interlace})}

\begin{proof}[Proof of Lemma~\ref{lem:recursion}]
\emph{Step 1: pivots are inverse corner elements.} Let $A_{[1:k]}$ be the leading $k\times k$ block of $A$. By Gaussian
elimination from the top, the $k$th pivot is the Schur complement of $A_{[1:k-1]}$ in $A_{[1:k]}$. For a tridiagonal matrix
only $b_{k-1}$ couples row $k$ to the block, so
\begin{equation}
L_k=a_k-b_{k-1}^2\,\big[A_{[1:k-1]}^{-1}\big]_{k-1,k-1}.
\end{equation}
Applying the block-inverse formula to $A_{[1:k]}$ shows $\big[A_{[1:k]}^{-1}\big]_{kk}=1/L_k$. Induction on $k$ then gives
the recursion $L_k=a_k-b_{k-1}^2/L_{k-1}$ with $[A_{[1:k-1]}^{-1}]_{k-1,k-1}=1/L_{k-1}$. The same argument from the bottom
gives $U_k=a_k-b_k^2/U_{k+1}$ and $[A_{[k+1:N_g]}^{-1}]_{k+1,k+1}=1/U_{k+1}$.

\emph{Step 2: diagonal element.} Partition $A$ around row $i$ into the top block, the $i$th row and the bottom block. The
only couplings of row $i$ are $b_{i-1}$ to the last row of the top block and $b_i$ to the first row of the bottom block. The
$(i,i)$ element of $A^{-1}$ is the inverse of the Schur complement of the two outer blocks:
\begin{align}
[A^{-1}]_{ii}
&=\Big[a_i-\frac{b_{i-1}^2}{L_{i-1}}-\frac{b_i^2}{U_{i+1}}\Big]^{-1}\nonumber\\
&=\frac{1}{L_i+U_i-a_i},
\end{align}
where the last step substitutes $L_i-a_i=-b_{i-1}^2/L_{i-1}$ and $U_i-a_i=-b_i^2/U_{i+1}$. See also
Ref.~\cite{Meurant1992}.
\end{proof}

\begin{proof}[Proof of Lemma~\ref{lem:interlace}]
\emph{Step 1: variational form.} With $x=S^{-1/2}y$, the generalized problem becomes a standard one,
\begin{equation}
\lambda_\mathrm{min}(H,S)=\min_{x\neq0}\frac{x^{\mathsf T}Hx}{x^{\mathsf T}Sx}
=\lambda_\mathrm{min}\big(S^{-1/2}HS^{-1/2}\big),
\end{equation}
by the Rayleigh--Ritz (Courant--Fischer) principle. The same holds for $(H_W,S_W)$.

\emph{Step 2: restriction.} Let $E$ be the matrix whose columns embed the window basis functions in the full basis, so
that $H_W=E^{\mathsf T}HE$ and $S_W=E^{\mathsf T}SE$. For $y\neq0$ in the window space, $x=Ey$ is a nonzero vector (columns of
$E$ are linearly independent), and
\begin{equation}
\frac{y^{\mathsf T}H_Wy}{y^{\mathsf T}S_Wy}=\frac{x^{\mathsf T}Hx}{x^{\mathsf T}Sx}.
\end{equation}

\emph{Step 3: conclusion.} The window minimum is therefore taken over a subset of the vectors used for the full minimum, so
\begin{equation}
\lambda_\mathrm{min}(H_W,S_W)\ge\lambda_\mathrm{min}(H,S).
\end{equation}
\end{proof}

\section{Exchange energy of the 1D soft-Coulomb gas}\label{app:ex}

We consider the uniform gas of density $n$ (spin-unpolarized) with interaction $w(r)=1/\sqrt{r^2+1}$.

\emph{Step 1: density matrix.} The occupied plane waves fill $|k|\le k_F$, and the spin-summed density is
$n=2\cdot2k_F/(2\pi)$, so $k_F=\pi n/2$. The spin-summed one-body density matrix is
\begin{equation}
\gamma(r)=2\int_{-k_F}^{k_F}\frac{dk}{2\pi}\,\ee^{\ii kr}=\frac{2\sin(k_Fr)}{\pi r}.
\end{equation}

\emph{Step 2: exchange energy as an integral.} The exchange energy per unit length is
$e_x=-\frac14\int_{-\infty}^{\infty}w(r)\,\gamma(r)^2\,dr$. Using $\gamma^2=4\sin^2(k_Fr)/(\pi^2r^2)$ and the evenness of the integrand,
\begin{align}
e_x&=-\frac{2}{\pi^2}\int_0^\infty\frac{\sin^2(k_Fr)}{r^{2}\sqrt{1+r^2}}\,dr\nonumber\\
&=-\frac{1}{\pi^2}\,I(2k_F),
\end{align}
where we used $\sin^2(k_Fr)=\tfrac12[1-\cos(2k_Fr)]$ and defined
\begin{equation}
I(b)=\int_0^\infty\frac{1-\cos br}{r^2\sqrt{1+r^2}}\,dr .
\end{equation}

\emph{Step 3: differential equation for $I$.} At $b=0$ the integrand vanishes, so $I(0)=0$. Differentiating under the
integral (the integrals converge uniformly),
\begin{align}
I'(b)&=\int_0^\infty\frac{\sin br}{r\sqrt{1+r^2}}\,dr,&I'(0)&=0,\\
I''(b)&=\int_0^\infty\frac{\cos br}{\sqrt{1+r^2}}\,dr=K_0(b),
\end{align}
where the last integral is a standard representation of the modified Bessel function $K_0$.

\emph{Step 4: integration.} Integrating $I''=K_0$ with $I(0)=I'(0)=0$, and then integrating by parts,
\begin{align}
I'(b)&=\int_0^bK_0(t)\,dt,\nonumber\\
I(b)&=\int_0^b\!\!\int_0^sK_0(t)\,dt\,ds=\int_0^b(b-t)K_0(t)\,dt\nonumber\\
&=b\int_0^bK_0(t)\,dt-\int_0^btK_0(t)\,dt\nonumber\\
&=b\int_0^bK_0(t)\,dt+bK_1(b)-1 .
\end{align}
The last line uses $\frac{d}{dt}[tK_1(t)]=-tK_0(t)$ and $tK_1(t)\to1$ as $t\to0$, so
$-\int_0^btK_0\,dt=bK_1(b)-1$.

\emph{Step 5: result.} With $2k_F=\pi n$,
\begin{align}
e_x(n)&=-\frac{1}{\pi^2}I(\pi n),\nonumber\\
v_x&=\frac{de_x}{dn}=-\frac{1}{\pi}I'(\pi n)\nonumber\\
&=-\frac1\pi\int_0^{\pi n}K_0(t)\,dt .
\end{align}
Equivalently, $e_x$ is a one-dimensional integral over the Fourier transform of $w$, which is proportional to $K_0$; the
steps above evaluate it in closed form. Parametrized local-density approximations for this gas, with correlation, are given in
Refs.~\cite{Helbig2011,Wagner2012}; we use the exchange part only.

\bibliography{refs}

\end{document}